\documentclass[a4paper,11pt]{iopart}

\usepackage{graphicx}
\usepackage{cite}

\renewcommand{\Re}{{\rm Re}}

\newcommand{\Ne}{\ensuremath{N_{\rm e}}}
\newcommand{\MI}{\ensuremath{M^{(1)}}}
\newcommand{\MII}{\ensuremath{M^{(2)}}}
\newcommand{\lrI}{\ensuremath{\lambda_r^{(1)}}}
\newcommand{\lqII}{\ensuremath{\lambda_q^{(2)}}}
\newcommand{\lrII}{\ensuremath{\lambda_r^{(2)}}}
\newcommand{\lup}{\ensuremath{\lambda_{+}}}
\newcommand{\lpd}{\ensuremath{\lambda_{+}^{*}}}
\newcommand{\lpr}{\ensuremath{\lambda_{+}^{\#}}}

\newcommand{\implies}{\Longrightarrow}

\begin{document}

\title[Voter models as a single-coordinate diffusion]{Ordering in voter models on networks: Exact reduction to a single-coordinate diffusion}
\author{R.\ A.\ Blythe}
\address{SUPA, School of Physics and Astronomy, University of Edinburgh, Mayfield Road, Edinburgh EH9 3JZ, UK\\[2ex]\today}

\begin{abstract}
We study the voter model and related random-copying processes on arbitrarily complex network structures.  Through a representation of the dynamics as a particle reaction process, we show that a quantity measuring the degree of order in a finite system is, under certain conditions, exactly governed by a universal diffusion equation.  Whenever this reduction occurs, the details of the network structure and random-copying process affect only a single parameter in the diffusion equation.  The validity of the reduction can be established with considerably less information than one might expect: it suffices to know just two characteristic timescales within the dynamics of a single pair of reacting particles.  We develop methods to identify these timescales, and apply them to  deterministic and random network structures. We focus in particular on how the ordering time is affected by degree correlations, since such effects are hard to access by existing theoretical approaches.
\end{abstract}

\section{Introduction}

The voter model \cite{cli73,hol75,cox86,lig99} is arguably the simplest nonequilibrium ordering process.  Its dynamics are perhaps most easily described within its original formulation as a model for population dynamics \cite{cli73}.  Each site of a lattice (or network) is occupied by an individual belonging to one of a number of species.  In each timestep, a site is chosen at random, and the individual at that site is replaced by an offspring (copy) of an individual selected at random from a neighbouring site.  In a finite system, one of the possible absorbing states---in which only a single species remains---is eventually reached.  From a physical perspective, the interest in this process is that order is arrived at purely through fluctuations, rather than driven by surface tension as in the Ising model (see e.g., \cite{dro99,dor01}; also \cite{cas09} for a recent review).

Many variants of the basic voter model have been investigated, some quite recently: examples include those where multiple individuals occupy a site \cite{bar09}, where the site chosen in each timestep contains the individual that is copied rather than replaced \cite{cas05aip,soo08}, or where different sites have different probabilities of being chosen for update \cite{mas10}.  Specific instances drawn from this wider class of \emph{random-copying} models have applications in population genetics, ecology and cultural evolution, as reviewed in \cite{bly07}.  Despite their simplicity, these models reproduce observed patterns of species diversity in ecosystems \cite{hub01} and of the cultural dynamics of features like baby names \cite{hah03}.  They also play an important role as \emph{null models} of evolution in the absence of selection \cite{hru09}.

A stochastic equation of motion for the frequency $\xi$ of a chosen species is straightforwardly obtained in the mean-field limit (i.e., when the system lacks spatial structure).  More precisely, when each site is visited for update at a uniform rate $m$, and the individual landing on the target site is sampled uniformly from the entire population, the probability distribution of $\xi$ is---for sufficiently large systems---governed by
\begin{equation}
\label{demf}
\frac{\partial}{\partial t}  P(\xi, t) = \frac{m}{\Ne} \frac{\partial^2}{\partial \xi^2} \xi(1-\xi) P(\xi, t) \;.
\end{equation}
Derivations of this equation can be found in many places, e.g., \cite{cro70,sla03,bly07}.  These stochastic dynamics have three key properties. First, there is no deterministic (drift) term in the above equation, confirming that order is attained through fluctuations alone.  Second, all timescales of the ordering process are proportional to the parameter $\Ne$: in the mean field, this parameter is simply equal to the number of sites, $N$.  Finally, the full, time-dependent solution for $P(\xi, t)$ from arbitrary initial condition is known \cite{cro70}.  Then, for example, the probability that complete order has been attained by time $t$ is given by the combination $P(0,t)+P(1,t)$.

Many recent works have switched focus to models defined on nontrivial network structures, notably \emph{heterogeneous} networks in which the degree can vary widely between nodes \cite{cas03,vil04,suc05epl,suc05pre,cas05,cas05aip,soo05,soo08,bax08,vaz08,sch09,pug09}.  Such networks are particularly relevant to cultural applications, where the interaction frequency between agents need not be purely a function of spatial proximity.  Remarkably, it has been found that even in the presence of significant heterogeneity, the ordering dynamics is well-described by the mean-field equation (\ref{demf}).  For example, the mean time to reach complete order as predicted by (\ref{demf}) has been found to apply in models with complex spatial structure \cite{soo05,soo08,bax08,vaz08,sch09,pug09}. The only differences are: (i) that $\xi$ is a weighted average of a species frequency over the network; and (ii) the \emph{effective size}, $\Ne$, need not equal the actual network size $N$ (or even scale linearly with it).  Thus, a good understanding of ordering behaviour within random-copying models on networks can be gained if one can identify when a reduction to (\ref{demf}) is possible, and if so, what the value of $\Ne$ is.  For example, the entire distribution of ordering times can be read off from the exact solution of (\ref{demf}).

In this work we will demonstrate that the mean-field diffusion equation (\ref{demf}) is in fact \emph{exact} for general random-copying processes on arbitrary networks at all but the very earliest times when there is a separation of emergent timescales within a two-particle coalescence reaction dynamics that is \emph{dual} to the random-copying process \cite{cox89,not90}.  In several earlier works \cite{soo05,cas05aip,soo08,bax08,vaz08,sch09,pug09}, the equation (\ref{demf}) was typically obtained by making specific assumptions and approximations, and the fundamental physical reason why the simple diffusion (\ref{demf}) should apply was not clearly identified.

The central object in our analysis is the \emph{quasi-stationary} distribution of a pair of particles that have survived for a long time without reacting.    This state decays at some characteristic rate, and if its lifetime greatly exceeds the relaxation time to the quasi-stationary state from a given initial condition, the model exhibits the separation of timescales that is required for the validity of (\ref{demf}).  The effective size $\Ne$ is then proportional to the single, dominant timescale in the reaction dynamics, that is, the lifetime of the quasi-stationary state.  

This paper is divided into two parts.  In the first part (Sections~\ref{modsec}--\ref{redsec}) we define the general class of random-copying models that includes well-studied instances, like the voter model, as special cases, and derive our main results.  These are: Eq.~(\ref{demf}), the condition for its validity and the expression for the fundamental parameter $\Ne$.  As we will need to switch repeatedly between the forward-time formulation of the dynamics in terms of a diffusion equation and an equivalent backward-time particle reaction process whose details are not widely discussed in the physics literature, we must unfortunately contend with lengthy, but necessary, preliminaries in Sections~\ref{modsec}--\ref{coalsec}.  Once this is done, the derivations of our main results follow in Section~\ref{redsec}.  We emphasise that the validity of Eq.~(\ref{demf}) can be determined by considering only a single pair of coalescing random walkers on the network.  In principle, ascertaining whether the system is completely ordered requires knowledge of the state of all $N$ sites simultaneously: hence being able to determine whether the considerably simpler single-coordinate diffusion applies without having to solve the full dynamics is of great practical benefit.

In the second part of this paper, we go beyond purely formal statements about random-copying dynamics and focus on strategies to estimate the emergent timescales of the two-particle reaction process.  These methods are developed in Section~\ref{trssec} and used to demonstrate the validity of the diffusion (\ref{demf}) for specific combinations of network structures and copying dynamics in Section~\ref{appsec}.  Although these methods are restricted to random-copying processes that map onto time-reversal symmetric random walks, it turns out that the most widely-studied variations on voter-model dynamics \cite{suc05epl,cas05aip,soo08,sch09,pug09} fall into this category.  On the other hand, we do not (in contrast to some existing treatments \cite{soo05,cas05aip,soo08,vaz08,sch09,pug09}) invoke a coarse-graining over multiple sites of the network to obtain a diffusion equation.  Thus, we can use our analytical methods to investigate the effects of degree correlations on the effective size $\Ne$ that appears in Equation (\ref{demf}), which until now has been accessible only by computer simulation.

We remark that the idea that random-copying dynamics in the presence of spatial structure may be adequately represented by a mean-field model with some effective size $\Ne$ has a very long history in population genetics, dating back to the work of Wright from the 1930s \cite{wri31}.  Furthermore, various separation of timescales arguments have been employed as a means to identify an effective size \cite{nor97,wak99,wak01,wak04,sjo05,mat06}.  In this work, our main contribution to this line of enquiry is to highlight the two particular emergent timescales that are relevant in determining the applicability (or otherwise) of a single effective size: their importance does not appear to have been fully recognised before. We furthermore show that this separation exists even in cases where it has been argued that there will be more than one relevant timescale in the dynamics, and that a description in terms of a single effective size will not apply \cite{sjo05}.  We discuss these points in much more detail in the conclusion, Section~\ref{consec}, along with a number of remaining open questions.

%% ----------

\section{Definition of a class of spatially-structured random-copying models}
\label{modsec}

\begin{figure}
\begin{center}
\includegraphics[width=0.33\linewidth]{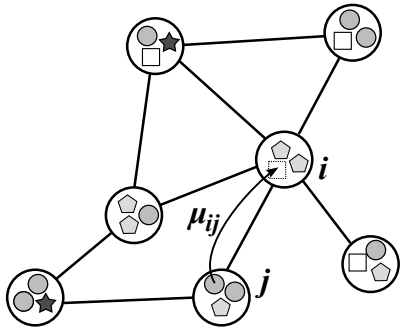}
\end{center}
\caption{\label{randomcopy} Random-copying dynamics on a network. Here, each site hosts $M=3$ individuals and there are four species in total.  The individual with dotted outline on site $i$ has been selected for replacement by copying an individual on site $j$.  This event takes place as a Poisson process with rate $\mu_{ij}$.}
\end{figure}

We consider a network of $N$ nodes (or sites) labelled by the integers $i=1,2,3,\ldots, N$.  Site $i$ hosts a population of $M$ individuals, each of which is a member of one of a number of species.  The dynamics is specified in terms of a set of copying rates $\mu_{ij}$.  Each corresponds to an independent Poisson process: at rate $\mu_{ij}$ one of the $M$ individuals on site $j$ is copied, and the offspring individual replaces a randomly-chosen individual on site $i$.  See Figure~\ref{randomcopy}. Note that the case $i=j$ is legitimate, and indeed the rates $\mu_{ii}$ will typically be nonzero in any particular instance of this model.  We could also consider models where $M$ varies from site to site within the framework presented below: however, this serves only to complicate certain definitions and expressions so we shall restrict the discussion to a spatially-uniform $M$.

The rates $\mu_{ij}$ determine the model's network structure.  Elements of the \emph{adjacency matrix} $A_{ij}$, which indicate if sites $i$ and $j$ are neighbours, are zero if $\mu_{ij}=0$, and unity otherwise (except the diagonal elements $A_{ii}$ which are always taken to be zero).  The \emph{degree} $z_i$ of node $i$ can then be defined as
\begin{equation}
z_i = \sum_{j} A_{ij} \;.
\end{equation}
Although we no not require the adjacency matrix to be symmetric in the first part of this paper, we will for simplicity assume this is the case. (For example, we do not then have to distinguish between in-degree and out-degree). Note, however, that the symmetry $A_{ij}=A_{ji}$ does \emph{not} imply that $\mu_{ij}=\mu_{ji}$.

In a finite system, there is some probability that a species eventually goes extinct---this occurs if all its representatives have been replaced by individuals of other species.  Over time, the number of species decreases until only a single one remains.  This stationary absorbing state of complete order is referred to variously as \emph{consensus} or \emph{fixation} (in the sociodynamics and genetics literature respectively).

\subsection{Voter, invasion and link dynamics}
\label{vilsec}

The original voter model dynamics specifies that, in each timestep, a site is first chosen at random, and then a copy taken from one of its neighbours.  This constrains the copying rates $\mu_{ij}$ to have the form
\begin{equation}
\label{muV}
\mu_{ij}^{V} = \frac{\mu A_{ij}}{z_i}
\end{equation}
where $\mu$ is the rate at which a site is chosen for update.  This is, however, only one of a vast range of possible ways to construct a set of copying rates.  Also discussed in the literature is the \emph{invasion process}, whereby the parent site $j$ is chosen uniformly at random in each timestep, and the offspring `invades' a randomly chosen neighbour (see, e.g., \cite{cas05aip,soo08}).  This corresponds to the choice
\begin{equation}
\label{muI}
\mu_{ij}^{I} = \frac{\mu A_{ij}}{z_j} \;.
\end{equation}
Another possibility is \emph{link dynamics} \cite{cas05aip,soo08}, in which an edge between two nodes is chosen at random, and the copy made in one of the two possible directions along that edge.  One way to specify this process is through the rates
\begin{equation}
\label{muL}
\mu_{ij}^{L} = \frac{\mu A_{ij}}{\bar{z}}
\end{equation}
where $\bar{z}$ is the mean degree within the network.  In networks where all nodes have the same degree, $z_i=z_j=\bar{z}$ for all $i,j$, and these three choices are all equivalent.  On heterogeneous networks, these different rules have been found to lead to different scalings of the effective size $\Ne$ with the actual network size $N$.  For concreteness, we will mostly consider the above copying schemes.  However, we stress that many of our results are quite general, and are not restricted to these specific choices.

\subsection{Discrete- and continuous-frequency models}
\label{dcfsec}

In the original formulation of the voter model, there is one individual per site: $M=1$.  Then, the frequency of a given species on a site can take only one of two values: $0$ or $1$.  To obtain a diffusion equation for the copying dynamics without the need to average over multiple sites (for example, all sites of the same degree, as has been done elsewhere \cite{soo05,cas05aip,soo08,vaz08,sch09,pug09}), we will take the limit $M\to\infty$, so that the frequency $x_i$ of a chosen species on a site becomes a continuous variable in the range $[0,1]$.  We will use the terminology \emph{discrete-frequency} and \emph{continuous-frequency} models to refer to the extreme cases $M=1$ and $M\to\infty$ respectively.

It is not \textit{a priori} obvious that results obtained for continuous-frequency models will apply to their discrete-frequency counterparts.  However, it turns out that for any fixed, finite $M$, discrete-frequency dynamics are recovered in a limit where the copying rates between sites, $\mu_{ij}$ where $i\ne j$, vanish.  This fact can be deduced from a convergence theorem for Markov chains due to M\"{o}hle \cite{moh98}.  The basic physical idea behind this result is that in the time it takes for an individual to arrive from a neighbouring site, sufficiently many replacements have been made by sampling from the local population of $M$ individuals that each site is in one of its locally absorbing states---that is, each site hosts only a single species.  As a consequence, the frequency of each species on any chosen site is always close to $0$ or $1$ (although globally it may be somewhere in between) and thus the discrete-frequency dynamics is recovered.

The way in which the $M=1$ discrete-frequency dynamics is obtained mathematically is summarised in \ref{conv}.  We illustrate here with the concrete examples of the previous subsection.  Discrete-frequency versions of voter, invasion or link dynamics can be obtained by taking the parameter $\mu$ that appears in (\ref{muV})--(\ref{muL}) to zero, whilst holding all $\mu_{ii}$ at fixed, finite (but otherwise arbitrary) values.   To recover the $M=1$ version of the voter, invasion or link dynamics, with the unit of time defined such that each site is visited on average once per timestep, one must multiply all times by $\mu / M$ (see \ref{conv}).

A key point is that for any finite $M$, no matter how large, the discrete-frequency dynamics are recovered for sufficiently small $\mu$.  That is, any difference between a set of frequencies $\{ x_i \}$ in the $M>1$ model and their values in corresponding realisations of the $M=1$ model can be made arbitrarily small by reducing $\mu$.  In the next section, and the remainder of this work, we will be working in the $M\to\infty$ limit, such that all frequencies $x_i$ can be regarded as continuous variables. We will assume that results for discrete frequency models can then be obtained by subsequently taking $\mu\to0$.  In other words, we assume that the limits $M\to\infty$ and $\mu\to0$ commute. We do not know \emph{a priori} that this assumption is valid. In practice, we find that we recover known results for the voter model, or good agreement with simulation, when we perform this procedure (see Section~\ref{appsec}).

\section{Forward-time formulation as a Fokker-Planck equation}
\label{fpesec}

We now derive the Fokker-Planck equation for probability distribution over the $N$-dimensional space of local frequencies $x_i$ of a chosen species.  Although the method used to derive this equation, based on \emph{jump moments}, is standard \cite{ris89}, we will recapitulate it here as it central to the derivation Eq.~(\ref{demf}) that follows in Section~\ref{redsec} below.  The key point is that the Fokker-Planck equation has the expansion
\begin{eqnarray}
\label{km}
\frac{\partial P(\{ x_i \}, \tau)}{\partial \tau} &=& 
- \sum_{i} \frac{\partial}{\partial x_i} a_i^{(1)}(\{ x_i \}, \tau) P(\{ x_i \}, \tau) +  {} \\\nonumber
&&\hspace{12ex} \frac{1}{2} \sum_{ij} \frac{\partial^2}{\partial x_i \partial x_j} a_{ij}^{(2)}(\{ x_i \}, \tau) P(\{ x_i \}, \tau) + \cdots
\end{eqnarray}
where $\tau$ is the time variable and the jump moments are defined as
\begin{equation}
\label{jm}
\fl a_{i_1 i_2 \cdots i_n}^{(n)}(\{ x_i \}, \tau) = \lim_{\delta \tau \to 0} \frac{ \langle [ x_{i_1}(\tau+\delta \tau) - x_{i_1}(\tau) ] \cdots [ x_{i_n}(\tau+\delta \tau) - x_{i_n}(\tau) ] \rangle}{\delta \tau} \;.
\end{equation}
In calculating the jump moments, the initial coordinates $x_i(\tau)$ are assumed known and fixed, and the angle brackets denote an average over all possible realisations of the dynamics in the time interval $[\tau, \tau+\delta \tau]$.  These averages can be computed from the underlying stochastic dynamical rules.

Within the present class of random-copying models, a local frequency $x_i$ may increase or decrease by an amount $\frac{1}{M}$ in any single update. The continuous-time transition rates for these processes are
\begin{eqnarray}
\label{Wup}
W\left(x_i \to x_i + \frac{1}{M}\right) &=& (1-x_i) \sum_{j} \mu_{ij} x_j \\
W\left(x_i \to x_i - \frac{1}{M}\right) &=& x_i \sum_{j} \mu_{ij} (1 - x_j) \;.
\label{Wdown}
\end{eqnarray}
In the time interval $[\tau, \delta \tau]$, only events involving a single change to one of the $x_i$ contributes to order $\delta \tau$ in the average in (\ref{jm}), for any value of $n$.  Thus, for the first jump moment we have,
\begin{equation}
\label{a1}
a_i^{(1)} =  \frac{1}{M} \sum_{j \ne i} \mu_{ij} (x_j - x_i) \;,
\end{equation}
and for the second moments
\begin{equation}
\label{a2}
a_{ii}^{(2)} = \frac{1}{M^2} \sum_j \mu_{ij} \left[ x_i(1-x_j) + x_j(1-x_i) \right]
\end{equation}
while $a_{ij}^{(2)} = 0$ for $i \ne j$, since two different $x_i$ never change at the same time.

If all $\mu_{ij}$ remain fixed as $M\to\infty$, the second jump moments vanish relative to the first.  The resulting Fokker-Planck equation then has only a drift (deterministic) term, and no diffusive component.  Both terms are contribute in this limit if one takes  $\mu_{ij}$ to vanish as $1/M$ for all $j\ne i$, as is standard in population genetics \cite{cro70,bly07}.  We thus define rescaled rates $m_{ij}$ and $c_i$ through
\begin{equation}
\label{muscale}
\mu_{ij} = \left\{ \begin{array}{ll} \displaystyle \frac{m_{ij}}{M} & i\ne j\\[2ex]
\displaystyle \frac{c_i}{2} & i = j \end{array} \right. \;.
\end{equation}
By introducing a rescaled time $t = \tau/M^2$ we find as $M\to\infty$ the Fokker-Planck equation
\begin{eqnarray}
\label{fpe}
\frac{\partial}{\partial t} P(\{ x_i \}, t) &=& \sum_{\langle i,j\rangle} \left(m_{ij} \frac{\partial}{\partial x_i} - m_{ji} \frac{\partial}{\partial x_i} \right) \left( x_i - x_j \right) P(\{ x_i \}, t) + {}
\nonumber\\
 &&\qquad \frac{1}{2} \sum_i c_i \frac{\partial^2}{\partial x_i^2} x_i(1-x_i) P(\{ x_i \}, t) \;.
\end{eqnarray}
The third and higher jump moments are all of order $1/M^3$ and higher, and thus do not contribute in the $M\to\infty$ limit.  Although this Fokker-Planck equation is strictly valid in this limit, it is understood that it serves as an excellent approximation for large but finite $M$ \cite{cro70}.  Our experience suggests that $M\sim100$ yields results consistent with infinite $M$ (see Section~\ref{bipsec} for a comparison between finite $M$ simulations and the infinite-$M$ theory).

In principle, to study the ordering dynamics of random-copying processes one would need to solve the full time-dependence of this $N$-dimensional diffusion.  It is therefore apparent that the reduction to the single coordinate diffusion (\ref{demf}), when it is possible, represents a dramatic reduction in the complexity of the problem.

\section{Backward-time formulation as a particle reaction process}
\label{coalsec}

\begin{figure}
\begin{center}
\includegraphics[width=0.33\linewidth]{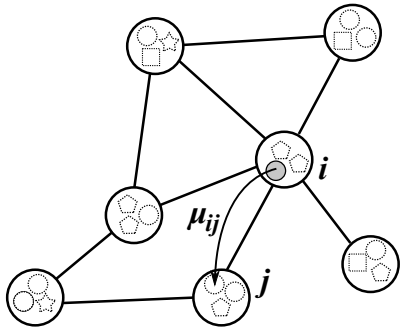}
\end{center}
\caption{\label{randomwalk} Backward-time formulation of the random-copying dynamics as a random walk process.  Shown is the configuration of the system of Figure~\ref{randomcopy} after the copying event has occurred.  Looking backward in time, the tagged individual on site $i$, shown with a solid fill, has a probability $\mu_{ij}$ per unit time of having been copied from site $j$.  The tagged individual thus hops to site $j$ at rate $\mu_{ij}$.}
\end{figure}

As mentioned in the introduction, the copying process defined in Section~\ref{modsec} has an equivalent formulation as a particle reaction dynamics (see e.g. \cite{cox89,not90} for mathematical formulations and \cite{bly07} for a more detailed account from a statistical physics perspective than the brief outline we give here).  This process is identified by looking backward in time at the history of an individual in the current configuration.  At some point in the past, this individual was created by the copying process.  The rate at which an individual currently on site $i$ was created by copying an individual on site $j$ is $\mu_{ij}/M$, where the factor $M$ appears because $\mu_{ij}$ is the total rate at which one of the $M$ individuals on site $i$ is replaced by an offspring arriving from site $j$.  This individual we call the \emph{ancestor} of that in the current configuration.  As we look further back in time, the ancestor performs a random walk, hopping from site $i$ to site $j$ at rate $\mu_{ij}$.  See Figure~\ref{randomwalk}.  We will also refer to these ancestors as \emph{particles}.

One can also consider the ancestry of multiple individuals.  These independently perform random walks, as just described.  However there is also the possibility that at some point in the history, one ancestor is the direct offspring of one of the others.  The rate at which one of a pair of ancestors occupying sites $i$ and $j$ was copied from the other is $(\mu_{ij} + \mu_{ji})/M^2$.  This time, the factor $M^2$ appears because both parent and offspring have a probability of $1/M$ of being involved in the copying process.  After this copying event, the two lines of ancestry merge: one individual is the common ancestor of the original pair.  This interaction can be viewed as a particle coalescence reaction.

To convert these rates into the system of rescaled time units defined in the previous section, we multiply them by $M^2$.  We then find that an ancestor hops from site $i$ to $j$ at rate $m_{ij}$, and a pair of particles on site $i$ react at rate $c_i$.  The rate for particles on different sites $i\ne j$ to react vanishes under this rescaling.

Within this backward-time picture, it is important to realise that if one starts with an arbitrary pair of individuals in a finite system and looks far enough back in time  the ancestor of one individual will at some point be the direct descendant of the other.  That is, any two particles eventually react, leading to an overall reduction in the number of particles over time (i.e., fewer ancestors who contribute descendants to the present-day population).  The steady state of the particle reaction process thus comprises a single common ancestor of any set of individuals chosen from the present-time population.  If there is a single ancestor common to the \emph{whole} population, then it follows that all individuals at the present time are of the same species---i.e., that complete order has been reached at the present time.  Thus the stationary absorbing state of full order in the forward-time dynamics corresponds directly to the stationary state of a single particle in the backward-time picture.

In the remainder of this work we need only to consider the backward-time dynamics starting from a single pair of particles. Two key properties of this reaction-diffusion process are the probability distributions $Q_k(t)$ that a single particle on site $i$ at time $t=0$ occupies site $k$ at time $t$ and $Q_{k\ell}(t)$ that a pair of particles occupy sites $k$ and $\ell$.  Note that in the reaction dynamics, we take increasing time to mean further back into the past, with $t=0$ indicating the present time.  When talking about a particle pair, we will consider the particles to be distinguishable, so that the configuration $k\ell$ is not the same as $\ell k$ if $k\ne\ell$.  These conventions make writing down the master equations for these probability distributions slightly more straightforward.

The master equation governing the single-particle distribution function is
\begin{equation}
\frac{\rm d}{{\rm d} t} Q_k(t) = \sum_{i \ne k} \left[ Q_i(t) m_{ik} - Q_k(t) m_{ki} \right]  \equiv - \sum_i Q_i(t) \MI_{ik} \;.
\end{equation}
Here we have introduced the matrix
\begin{equation}
\label{MI}
\MI_{ik} = \left\{ \begin{array}{ll} -m_{ik} & i \ne k \\ \sum_j m_{ij} & i = k \end{array} \right. \;,
\end{equation}
where the sign of the entries has been chosen so that the eigenvalues of $\MI$ all have nonnegative real parts.  Assuming that the network is such that the stationary distribution $Q_k = \lim_{t\to\infty} Q_k(t)$ is unique, we have at late times
\begin{equation}
Q_k(t) \sim Q_k + a u_k {\rm e}^{-\lambda_r^{(1)} t}
\end{equation}
where $\lambda_r^{(1)}$ is the eigenvalue of $\MI$ with the smallest, nonzero real part.

Meanwhile, the joint probability distribution for a pair of unreacted particles is governed by the master equation
\begin{equation}
\frac{{\rm d}}{{\rm d} t} Q_{k\ell}(t) = \sum_{i} Q_{i\ell}(t) m_{ik} + \sum_{j} Q_{kj}(t) m_{j\ell} - Q_{kk}(t) c_k \delta_{k\ell} \;.
\end{equation}
The first two terms in this equation give the contributions to this probability from particles hopping into, or out of, sites $k$ and $\ell$.  The final term arises from the decrease in the probability that the two particles have not yet reacted due to the coalescence reaction. We can write this equation more compactly as
\begin{equation}
\frac{{\rm d}}{{\rm d} t} Q_{k\ell}(t) = - \sum_{ij} Q_{ij}(t) \MII_{ij,k\ell}
\end{equation}
where the matrix $\MII_{ij,k\ell}$ is
\begin{equation}
\label{MII}
\MII_{ij,k\ell} = - m_{ik} \delta_{j\ell} - m_{j\ell} \delta_{ik} + c_i \delta_{ijk\ell} \;.
\end{equation}
Here we use $\delta_{ijk\ell}$ as a shorthand for $\delta_{ij}\delta_{jk}\delta_{k\ell}$, i.e., the object that is zero unless all four indices are the same, in which is it equals unity.

On a finite network where the stationary distribution for a single particle is unique, the two particles will eventually react.  Therefore $Q_{k\ell}(t)$ vanishes as $t\to\infty$.  The matrix $\MII$ has eigenvalues whose real parts are all positive. The smallest eigenvalue $\lambda_q^{(2)}$ is strictly positive, and is equal to the asymptotic rate at which a pair of unreacted particles coalesce.  The associated eigenvector we will denote $Q_{k\ell}$ and normalise so that $\sum_{k\ell} Q_{k\ell}=1$.  This is the \emph{quasi-stationary distribution} of the two-particle system.  That is, if we condition on the two particles never having reacted, the probability they are found on sites $k$ and $\ell$ approach the stationary value $Q_{k\ell}$ \cite{dar67}.  The rate of relaxation to the quasi-stationary state is given by the real part of the second-smallest eigenvalue $\lambda_r^{(2)}$.  That is, at late times
\begin{equation}
\label{Q2decay}
Q_{k\ell}(t) \sim Q_{k\ell} {\rm e}^{-\lambda_q^{(2)} t}+ b v_{k\ell} {\rm e}^{-\lambda_r^{(2)} t} \;.
\end{equation}
As stated in the introduction, and to be demonstrated below, it is a separation of timescales between the relaxation to and decay of the quasi-stationary state, controlled by $\lambda_r^{(2)}$ and $\lambda_q^{(2)}$ respectively, that allows for the reduction of the full Fokker-Planck equation (\ref{fpe}) to the single-coordinate diffusion of Eq.~(\ref{demf}).

\section{Exact reduction to a single-coordinate diffusion equation}
\label{redsec}

Now that the forward- and backward-time formulations of the model defined in Section~\ref{modsec} have been set up, we may derive the single-coordinate diffusion (\ref{demf}) and establish the condition for its validity.  This amounts to identifying the quantity $\xi$ whose ensemble-average is conserved by the dynamics, and calculating its second jump moment.  The procedure is not hard, but it does involve several manipulations that require explanation.

The key quantity $\xi$ has been identified a number of times before (see e.g., \cite{suc05epl,cas05aip,soo08,bax08}).  It is 
\begin{equation}
\label{xi}
\xi(t) = \sum_i Q_i x_i(t) \;,
\end{equation}
where we recall that $Q_i$ is the stationary distribution of the single-particle random walk, and $x_i$ is the frequency of the species of interest on site $i$.  To show that the diffusion equation (\ref{demf}) has no deterministic term, we calculate the first jump moment
\begin{eqnarray}
a^{(1)}(\xi, t) &=& \sum_i Q_i \lim_{\delta t \to 0} \frac{\langle x_i(t+\delta t) - x_i(t) \rangle}{\delta t}\\
 &=& \sum_i Q_i \frac{{\rm d}}{{\rm d} t} \langle x_i(t) \rangle = \frac{{\rm d}}{{\rm d} t} \langle \xi(t) \rangle \;.
\end{eqnarray}
By multiplying (\ref{fpe}) by $x_i$ and integrating over all the $x$ coordinates (see \cite{bax06,bax08}), we find
\begin{equation}
\frac{{\rm d}}{{\rm d} t} \langle x_i(t) \rangle = \sum_j \left[ m_{ij} \langle x_j(t) \rangle - m_{ji} \langle x_i(t) \rangle \right] = - \sum_j \MI_{ij} \langle x_j(t) \rangle
\end{equation}
where $\MI$ is as defined in Eq.~(\ref{MI}). Hence,
\begin{equation}
\label{a1xi}
a^{(1)}(\xi, t) = - \sum_j \left( \sum_{i} Q_i \MI_{ij}\right) \langle x_j(t) \rangle  = 0
\end{equation}
where the second equality follows because $Q_i$ is the stationary distribution of the single-particle random walk dynamics.  In other words, it is the left eigenvector of the matrix $\MI$ with eigenvalue zero.  Thus the first jump moment of $\xi$ vanishes, as required.

We now turn to the second jump moment of $\xi$.  For brevity, we define $x_i \equiv x_i(t)$ and $x_i' \equiv x_i(t+\delta t)$.  Then,
\begin{eqnarray}
a^{(2)}(\xi, t) &=& \sum_{ij} Q_i Q_j \lim_{\delta t \to 0} \frac{ \langle [x_i' - x_i] [x_j' - x_j] \rangle}{\delta t} \\
&=& \sum_{ij} Q_i Q_j \lim_{\delta t \to 0} \frac{ \langle x_i' x_j' \rangle - x_i x_j}{\delta t}  - 2 \sum_{ij} Q_i Q_j \lim_{\delta t \to 0} \frac{ x_i ( \langle x_j' \rangle - x_j)}{\delta t} \\
&=& \sum_{ij} Q_i Q_j \frac{{\rm d}}{{\rm d} t} \langle x_i(t) x_j(t) \rangle - 2 \xi \frac{{\rm d}}{{\rm d} t} \langle \xi(t) \rangle \;.
\label{a2ds}
\end{eqnarray}
Notice that we take all the $x_i$ variables to have known values at time $t$. There is a slight complication here: only the value of $\xi$ is known at time $t$, not the individual $x_i$.  However, we will shortly find that at late times $a^{(2)}(\xi)$ depends only on $\xi$, and hence the resulting expression will apply no matter what the actual individual values of $x_i$ are in any given realisation of the dynamics.

As we have shown above, $\langle \xi(t) \rangle$ has a zero time derivative, so only the first term in the previous expression contributes to the second jump moment.  To make progress, it is useful to introduce the quantity
\begin{equation}
\label{Pdef}
P_{ij}(t) = \langle x_i(1-x_j) \rangle + \langle (1-x_i)x_j \rangle \;.
\end{equation}
This is the probability that two individuals randomly selected from sites $i$ and $j$ are of different species.  On a finite network, this quantity asymptotically decays to zero, and in fact, its time evolution is governed by the equation
\begin{equation}
\label{dPijdt}
\frac{{\rm d}}{{\rm d} t} P_{ij}(t) = - \MII_{ij,k\ell} P_{k\ell}(t) 
\end{equation}
where $\MII$ is the transition matrix for the two-particle reaction process defined by Eq.~(\ref{MII}).  One way to see this is to calculate the time derivative of $\langle x_i x_j \rangle$ by multiplying (\ref{fpe}) by $x_i x_j$ and integrating, as described previously.  A more direct way is to consider events that cause the probability that two individuals are of different species to change in a time interval $[t,t+\delta t]$ (see, e.g., \cite{whi97}).  The rate at which $P_{ij}(t)$ changes due to a parent on site $k$ leaving an offspring on site $i$ is $m_{ik} [P_{kj}(t) - P_{ij}(t)]$.  Meanwhile, if both individuals are on the same site $i$, there is a probability $c_i$ per unit time that one of them replaces the other (see Section~\ref{coalsec}), an event that causes the probability that the two individuals are distinct to decrease at rate $c_i P_{ii}(t)$.  Summing over all events that change $P_{ij}$ we find
\begin{equation}
\frac{{\rm d}}{{\rm d} t} P_{ij}(t) = \sum_{k\ne i} m_{ik} [P_{kj}(t) - P_{ij}(t)] + \sum_{\ell \ne j} m_{j\ell} [P_{i\ell}(t) - P_{ij}(t)] - c_i \delta_{ij} P_{ii}(t) 
\end{equation}
which is equivalent to (\ref{dPijdt}).

Since the time evolution of $P_{ij}(t)$ and the particle pair distribution function $Q_{k\ell}(t)$ discussed in Section~\ref{coalsec} are both determined by the same matrix $\MII$, $P_{ij}(t)$ has a late-time expansion of the same general form as that given by Eq.~(\ref{Q2decay}) for $Q_{k\ell}(t)$.  That is,
\begin{equation}
\label{P2decay}
P_{ij}(t) \sim P_{ij} {\rm e}^{-\lambda_q^{(2)} t} + c w_{ij} {\rm e}^{-\lambda_r^{(2)} t}
\end{equation}
Now let us suppose there is some limit, for example $N \to \infty$, in which
\begin{equation}
\label{cond}
\frac{\lambda_q^{(2)}}{\Re \lambda_r^{(2)}} \to 0 \;.
\end{equation}
Then, under a further rescaling of time $t' = \lqII t$, we have in this limit at any $t'>0$
\begin{equation}
P_{ij}(t') = P_{ij} {\rm e}^{-t'} \quad\implies\quad \frac{{\rm d}}{{\rm d} t'} P_{ij}(t') = - P_{ij}(t') \;.
\end{equation}
In other words, at times $t'>0$, the system is assumed to be in the quasi-stationary state.  To avoid too many different time units, we shall remain in the unscaled time $t$ and assume that the quasi-stationarity condition (\ref{cond}) is satisfied, and thus that $\frac{{\rm d}}{{\rm d} t} P_{ij}(t) = - \lqII P_{ij}(t)$ when $t>0$.

To complete the derivation of the second jump moment $a^{(2)}(\xi)$, we differentiate both sides of (\ref{Pdef}) with respect to time $t$, rearrange and substitute into (\ref{a2ds}).  We find
\begin{eqnarray}
\label{a2xit1}
a^{(2)}(\xi, t) &=& \frac{{\rm d}}{{\rm d} t} \langle \xi(t) \rangle - \frac{1}{2} \sum_{ij} Q_i Q_j \frac{{\rm d}}{{\rm d} t} P_{ij}(t) \\
&=& \frac{\lqII}{2} \sum_{ij} Q_i Q_j P_{ij}(t) \\
&=& \frac{\lqII}{2} \sum_{ij} Q_i Q_j ( x_i + x_j - 2 x_i x_j) = \lqII \xi(1-\xi) \;.
\end{eqnarray}
To arrive at the second line we used the fact that the average of $\xi$ has zero time derivative, that (\ref{cond}) holds and $t>0$. The derivation is completed using the definition of $P_{ij}(t)$, Eq.~(\ref{Pdef}), and of $\xi$, Eq.~(\ref{xi}).  As advertised, this second jump moment is a function only of the collective coordinate $\xi$.  Using the expansion (\ref{km}) we find that the  motion of the collective coordinate $\xi$ is governed by the Fokker-Planck equation
\begin{equation}
\label{main}
\frac{\partial}{\partial t}  P(\xi, t) = \frac{\lqII}{2} \frac{\partial^2}{\partial \xi^2} \xi(1-\xi) P(\xi, t) \;.
\end{equation}
at any time $t>0$ in a limit where the condition (\ref{cond}) is satisfied.  This is Eq.~(\ref{demf}) with an effective size 
\begin{equation}
\Ne = \frac{2m}{\lqII}
\end{equation}
where we recall $m$ is the baseline rate of the copying process.  This concludes the derivation of the main result of this work.  In the remaining sections we focus on strategies for determining when the key condition (\ref{cond}) is satisfied, and for estimating the single parameter $\lqII$ (or equivalently, $\Ne$) that depends on the network structure and choice of copying dynamics when the diffusion (\ref{demf}) or (\ref{main}) is valid.

We conclude this section with a couple of remarks on the derivation of (\ref{main}).

\paragraph{Behaviour at time $t=0$} At very early times, and technically only at time $t=0$ in the limit where (\ref{cond}) holds, there are additional contributions to the second jump moment.  Whether looking forwards or backwards in time, their net effect is to project the initial condition onto the quasi-stationary state.  To be more precise, if the leading eigenvectors of the matrix $\MII$ are normalised so that
\begin{equation}
\sum_{ij} P_{ij} Q_{ij} = 1 \;,
\end{equation}
the probability that two particles initially occupying sites $i$ and $j$ have not reacted by time $t$ and occupy sites $k$ and $\ell$ is
\begin{equation}
D(ij|k\ell; t) = P_{ij} Q_{k\ell} {\rm e}^{-\lqII t}
\end{equation}
if all but the quasi-stationary modes are neglected at times $t>0$.  By taking the limit $t\to0$ from above, and summing over all $k$ and $\ell$, we find that $P_{ij}$ is the probability that this particle pair does not react before the quasi-stationary state is reached.

This immediate collapse onto the quasi-stationary state has implications on the initial condition for the diffusion equation (\ref{demf}), given an initial set of frequencies $x_i(0)$.  It is possible to calculate the variance in $\xi$ at the onset of quasi-stationarity by appealing to (\ref{Pdef}).  In the simple, solvable examples discussed in Section~\ref{appsec} below where we find (\ref{demf}) to be valid (Wright's island model and the bipartite network), we find that for any given initial combination of $x_i(0)$ a delta-function initial condition $P(\xi,0) = \delta(\xi - \sum_i Q_i x_i[0])$ is appropriate.  Whether this is generally the case is a question we leave open for future work.

\paragraph{Diffusion of other collective coordinates} It turns out that any collective coordinate of the form
\begin{equation}
\label{chi}
\chi(t) = \sum_i \omega_i x_i(t)
\end{equation}
where the $\omega_i$ are an arbitrary set of positive weights that sum to unity, satisfies the diffusion equation (\ref{demf}) as long as, in addition to (\ref{cond}) the condition
\begin{equation}
\label{cond2}
\frac{\lqII}{\Re\lrI} \to 0 
\end{equation}
is also satisfied.

This can be seen by repeating the above derivation of the jump moments with this more general set of weights.  Ultimately, one finds
\begin{eqnarray}
a^{(1)}(\chi, t) &=& \frac{{\rm d}}{{\rm d} t} \langle \chi(t) \rangle \\
a^{(2)}(\chi, t) &=& (1 - 2\chi) \frac{{\rm d}}{{\rm d} t} \langle \chi(t) \rangle - \frac{1}{2} \sum_{ij} \omega_i \omega_j \frac{{\rm d}}{{\rm d} t} P_{ij}(t) \;.
\end{eqnarray}

The leading decay modes of the first jump moment is governed by the eigenvalue $\lrI$, while the modes corresponding to $\lqII, \lrII$ and $\lrI$ contribute to the second.  Thus we require both (\ref{cond}) and (\ref{cond2}) in order to obtain a vanishing first jump moment, and a time-dependent second jump moment of the form
\begin{equation}
a^{(2)}(\chi, t) = \lqII \chi(1-\chi)
\end{equation}
at times $t>0$.

Although all collective coordinates of the form (\ref{chi}) have the same diffusion equation---and crucially, the same effective size $\Ne$---when these conditions hold, we prefer the definition (\ref{xi}) because its first jump moment vanishes at all times, not just in the quasi-stationary state, and only the two-particle decay modes contribute to the second, as opposed to a mixture of the one- and two-particle modes.  As a consequence, the condition for validity of the diffusion (\ref{xi}) is slightly weaker and cleaner, as it makes reference only to the two leading decay modes within the two-particle sector.  Furthermore, the effect of the dynamics at time $t=0$ before the onset of the quasi-stationary state is harder to predict when the first jump moment does not vanish identically.

\section{Methods of analysis for models with time-reversal symmetry}
\label{trssec}

So far in this work, we have shown that if there is a separation of timescales in random-copying models on networks between the rate of relaxation to the quasi-stationary state and the lifetime of that quasi-stationary state, the exactly solvable diffusion equation (\ref{demf}) correctly describes the ordering dynamics at all but the earliest times.  To demonstrate the validity of (\ref{demf}), and compute the crucial parameter $\Ne$, one must in principle find the two smallest eigenvalues of the two-particle reaction matrix $\MII$ and verify the condition (\ref{cond}).  Whilst this represents a major simplification of the full dynamics, there is of course no simple, general expression for these eigenvalues for an arbitrary set of transition rates $m_{ij}$ and $c_i$.

In this section, we describe simplifications that arise in models where the single-particle hopping process exhibits time-reversal symmetry (detailed balance).  In particular, this allows one to relate the relaxation rate to the two-particle quasi-stationary state, $\lrII$, to the relaxation rate to the single-particle stationary state $\lrI$.  This leads to a weaker, but more easily verified, condition for the validity of the diffusion (\ref{demf}).  Furthermore, the symmetry allows one to develop a systematic variational method to bound the asymptotic reaction rate $\lqII$ from above, or equivalently, the effective size $\Ne$ from below.  We develop these methods in this section, and apply them to specific combinations of network structure and update rules in the next.

\subsection{Time-reversal symmetry and its consequences}

The stationary state of the single-particle random walk is time-reversal symmetric if the detailed balance condition
\begin{equation}
\label{db1}
Q_i m_{ij} = Q_j m_{ji} \quad \forall i,j
\end{equation}
is satisfied.  Although this may seem an overly restrictive constraint on the rates $m_{ij}$, it applies, for example, if they have the general form $m_{ij} = m A_{ij} \alpha_i \beta_j$, where $A_{ij}$ is the adjacency matrix and $\alpha_i$ and $\beta_j$ are arbitrary positive site-dependent quantities.  The three sets of rules discussed in Section~\ref{vilsec} fall into this category, and so the methods we derive here will be applicable to these cases of interest.

The relation (\ref{db1}) implies a corresponding symmetry in the matrix $\MII$:
\begin{equation}
\label{db2}
Q_i Q_j \MII_{ij,k\ell} = Q_k Q_{\ell} \MII_{k\ell,ij} \;.
\end{equation}
As a consequence of this symmetry, all eigenvalues $\lambda$ of $\MII$ are real, and for any right eigenvector $\phi_{ij}^{\lambda}$ the corresponding left eigenvector is $\psi_{ij}^{\lambda} \propto Q_i Q_j \phi_{ij}^{\lambda}$ \cite{pea65}. The eigenvectors form an orthonormal set, i.e., 
\begin{equation}
\sum_{ij} \psi_{ij}^{\lambda} \phi_{ij}^{\lambda'} \propto \sum_{ij} Q_i Q_j \phi_{ij}^{\lambda} \phi_{ij}^{\lambda'} = 0 \quad\mbox{if $\lambda \ne \lambda'$} \;.
\end{equation}
In particular, this means that the key quantities $P_{ij}$ and $Q_{ij}$ characterising the quasi-stationary state are related by $Q_{ij} = A Q_i Q_j P_{ij}$, where the constant $A$ is fixed by the normalisation conditions $\sum_{ij} Q_{ij} = \sum_{ij} P_{ij} Q_{ij} = 1$.  These properties are crucial in the analysis of the quasi-stationary state that follows.

\subsection{Relationship between one- and two-particle relaxation rates}
\label{1vs2sec}

We now show that, as a consequence of (\ref{db2}), $\lrII \ge \lrI$, and hence that the condition (\ref{cond2}) implies the condition (\ref{cond}) on the validity of the diffusion equation (\ref{demf}).  The condition (\ref{cond2}) may be easier to demonstrate than (\ref{cond}), since the relaxation rate of the single-particle dynamics is in principle more accessible than that for the two-particle dynamics.  Indeed there are methods to bound $\lrI$ from above \cite{dia91} which would provide a means to demonstrate that (\ref{cond2}) is satisfied, although we will not use them here.

One way to show that $ \lrII \ge \lrI$ is to use standard perturbation theory \cite{pea65}.  We assume we have the full set of eigenvalues and eigenvectors of the matrix $\MII$ for some combination of $m_{ij}$ and $c_i$, and then increase the coalescence rate on site $x$ by an amount $\delta c$.  The perturbation of $\MII$ is then $\delta \MII_{ij,k\ell} = \delta c$ if $i=j=k=\ell=x$, and zero otherwise.  Then, for an eigenvalue $\lambda$ of $\MII$ we have from first order perturbation theory \cite{pea65}
\begin{equation}
\delta \lambda = \delta c \frac{\sum_{ijk\ell} \psi^{\lambda}_{ij}\, \delta \MII_{ij,k\ell}\, \phi^{\lambda}_{k\ell}}{\sum_{ij}\psi^{\lambda}_{ij}\phi^{\lambda}_{ij}} = \delta c \frac{Q_x^2 [\phi^{\lambda}_{xx}]^2}{\sum_{ij} Q_i Q_j [\phi^{\lambda}_{ij}]^2} \;.
\end{equation}
This is a nonnegative quantity. Hence, in particular, $\lrII$ must increase (or remain constant) as reaction rates $c_i$ are increased from zero.  When all $c_i=0$, the reactions have no effect, and the two particles independently evolve according to the single-particle process governed by $\MI$.  Hence, the second smallest eigenvalue of $\MII$ equals that of $\MI$ when all $c_i = 0$.  Since $\lrII$ cannot decrease as the reactions are turned on, we have $\lrII \ge \lrI$, as claimed.

\subsection{Variational approach for bounding the asymptotic coalescence rate}

Since the eigenvalue $\lqII$ that sets the characteristic timescale of the diffusion (\ref{demf}) is the smallest of a set of real eigenvalues, we can use a variational principle \cite{pea65} to bound it from above.  Let us denote such a bound $\lup$.  The utility of this is twofold.  First, if one can show that in some limit
\begin{equation}
\label{cond3}
\frac{\lup}{\lrII} \to 0 \quad\mbox{or}\quad \frac{\lup}{\lrI} \to 0 \;,
\end{equation}
then the condition (\ref{cond}) for the validity of (\ref{demf}) follows.  Second, the expression for $\lup$ can be used as an estimate of the asymptotic reaction rate $\lqII$.

One can construct a bound $\lup$ from an arbitrary vector $\phi_{ij}$ through the expression
\begin{equation}
\label{var}
\lqII \le \lup = \frac{\sum_{ijk\ell} Q_i Q_j \phi_{ij} \MII_{ij,k\ell} \phi_{k\ell}}{\sum_{ij} Q_i Q_j \phi_{ij}^2} \;,
\end{equation}
which is obtained by expanding the vector $\phi_{ij}$ as a sum over the eigenvectors of the matrix $\MII$ (see, e.g., \cite{pea65} for a derivation).

There are two key points here.  First, this result does not rely on being able to solve the two-particle reaction process.  Any choice of $\phi_{ij}$ leads to an expression for $\lup$.  Even if this choice departs dramatically from the exact quasi-stationary eigenvector $P_{ij}$, it may still be possible to show that one of the ratios (\ref{cond3}) vanishes, in turn assuring validity of (\ref{demf}).  The second point is that, in the usual way, one can include variational parameters in the expression for $\phi_{ij}$, and minimise $\lup$ with respect to these parameters to obtain the best estimate of $\lqII$ available within with the space of vectors spanned by them. We will encounter explicit examples in the next section, details of which are in the Appendix.

\section{Applications}
\label{appsec}

In order to make concrete the ideas and methods developed in the previous two sections, we now discuss their application to specific combinations of copying dynamics and network structures.  In addition to checking certain known limits, we will use the fact that our approach does not invoke a coarse-graining over multiple lattice sites to obtain new theoretical results for the effect of correlations on the effective size $\Ne$.

\subsection{Wright's island model}
\label{wimsec}

A spatially-structured random-copying model that is prominent in the population genetics literature is Wright's island model \cite{wri43}.  In this model, each site receives an offspring from one of the other sites (chosen uniformly) at a constant rate $m$.  That is, $m_{ij} = \frac{m}{N-1}$ for all $i \ne j$.  The coalescence rate is also taken to be uniform: $c_i = c$ for all $i$.  The limited spatial structure in this model admits exact calculation of the eigenvalues $\lrI$ and $\lqII$, and therewith demonstration that (\ref{demf}) is exact in the $N\to\infty$ limit.  Despite the venerability of Wright's island model, we believe that the validity of the diffusion equation (\ref{demf}) has not been shown explicitly before, although the resulting expression for the effective population size $\Ne$ is well known (see e.g.,~\cite{wak99}; also \cite{wan99} for a review).

We begin by examining the single-particle dynamics. Since all sites are equivalent, the stationary distribution is uniform: $Q_i = \frac{1}{N}$.  The rate of relaxation to this steady state, $\lrI$, is the smallest nonzero eigenvalue of the eigenvector $\MI_{ij} = \frac{m}{N-1} (N \delta_{ij} - 1)$.  One can verify by inspection that any $N$-dimensional vector will a single entry equal to $1$, another equal to $-1$ and the rest zero is an eigenvector of $\MI$.  Since there are $N-1$ linearly-independent vectors of this form, all $N-1$ nonzero eigenvalues are degenerate, and one finds
\begin{equation}
\lambda_r^{(1)} = m \frac{N}{N-1} \;.
\end{equation}

We now turn to the two-particle dynamics.  It is possible to calculate $\lqII$ exactly.  One way to do this is with the variational method described above.  We note that in the quasi-stationary state, all knowledge of the initial condition is lost.  Thus there are only two distinct two-particle configurations in the quasi-stationary state: one in which both particles are on the same site, and one in which they are on distinct sites.  To construct the ansatz, we set all diagonal entries $\phi_{ii}=\alpha$ and all off-diagonal entries equal to unity.  The overall normalisation of the vector is unimportant, since this cancels in the numerator and denominator of (\ref{var}).  Thus we can without loss of generality fix one of the variational parameters and optimise the bound (\ref{var}) with respect to those that remain.  Since the exact quasi-stationary eigenvector is obtained with a particular value of $\alpha$, it follows that by minimising the right-hand of (\ref{var}) with respect to $\alpha$ will give us the exact decay rate of the quasi-stationary state.

In the Appendix, we show that in the quasi-stationary state, $\alpha$ and $\lqII$ are given by the solution of the pair of equations
\begin{eqnarray}
\label{wimsim1}
\alpha &=& \frac{2m}{2m+c-\lqII} \\
\lqII &=& \frac{\alpha c}{N-1+\alpha} \;.
\label{wimsim2}
\end{eqnarray}
Although these equations admit two solutions for $\lqII$, only one is positive.  This is
\begin{equation}
\label{wim}
\lqII = \frac{2mN+c(N-1)}{2(N-1)} \left[ \sqrt{1 + \frac{8mc(N-1)}{[2mN+c(N-1)]^2}} -1 \right] \;.
\end{equation}

We can now use condition (\ref{cond2}) to identify limits in which the diffusion (\ref{demf}) with effective size $\Ne = 2m/\lqII$ is valid.  One is the \emph{fast-migration limit}, $\frac{c}{m}\to0$, for which exact results are available on arbitrary network structures \cite{nag80}.  For Wright's island model we have
\begin{equation}
\frac{\lqII}{\lrI} = \frac{N-1}{N^2} \frac{c}{m} + O\left(\left[\frac{c}{m}\right]^2\right) 
\end{equation}
indicating that in this limit, (\ref{demf}) is valid for any $N$.  In this regime, $\lqII \sim c/N$.  This is consistent with the general result obtained by other means \cite{nag80}, and thus serves as a basic check of the methods derived here.

For general $c$ and $m$, we have in the infinite system size limit, $N\to\infty$, that
\begin{equation}
\frac{\lqII}{\lrI} = \frac{2c}{2m+c} \frac{1}{N} + O \left(\frac{1}{N^2}\right) \;.
\end{equation}
Applying the condition (\ref{cond2}) we find that the diffusion (\ref{demf}) is valid for any combination of $m$ and $c$ with an effective size $\Ne = N (1 + 2 \frac{m}{c})$.  This is in agreement with the large-$N$ expression for the effective population size quoted in the literature (see, e.g., \cite{wak99,wan99}).  
To achieve this agreement, however, one needs to match up the choices of time units and copying rates that are used in standard population genetics models: the correct choice has the literature effective size related to our own by $\Ne^{\rm (lit)} = \Ne/4m$ and the reaction rate $c = 1/2M$ where $M$ is the subpopulation size. In these units, $m$ is the fraction of individuals replaced by immigrants in each generation.  What is perhaps more significant is that, as far as we are aware, the applicability of this single effective population size over extended timescales has not been demonstrated explicitly even for this extremely simple model, even though it is routinely assumed.

We remark that in the limit $m\to0$ where discrete-frequency models, like the original voter model, are recovered (see section~\ref{dcfsec}), we find that $\Ne = N$, as expected.  Thus at least in this case, the limits of infinite subpopulation size $M$ and vanishing migration rates $m$ indeed commute.

\subsection{Bipartite network}
\label{bipsec}

The bipartite network, previously examined in \cite{soo05,soo08,sch09}, is insightful because it is perhaps the simplest network structure for which the three rules of Section~\ref{vilsec} are not equivalent.  Furthermore, as we will see below, degree correlations can be relevant, in the sense that they can lead to departures from results previously obtained by assuming an absence of such correlations \cite{vil04,soo05,soo08,pug09}.

\begin{figure}
\begin{center}
\includegraphics[width=0.5\linewidth]{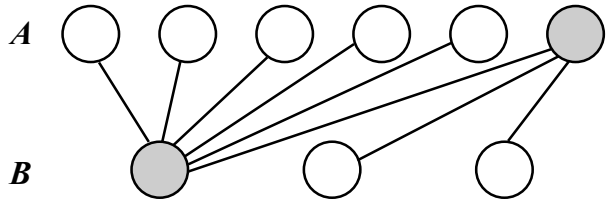}
\end{center}
\caption{\label{bipartite} Bipartite network comprising two groups of nodes, $A$ and $B$.  Each site in group $A$ may receive copies from any site in group $B$ with equal probability per unit time, and vice versa.  For clarity, the full set of edges has been shown only for the shaded nodes.}
\end{figure}

The $N$ nodes of the bipartite network are divided into two groups, $A$ and $B$, with a number $n_A$ of the nodes in group $A$, and $n_B$ in group $B$.  The rate $m_{ij}$ at which an individual $i$ from group $A$ is replaced from a copy of an individual $j$ from group $B$ is $m_{ij}=m_{AB}$.  At rate $m_{BA}$, copying takes place in the reverse direction. As in Wright's island model, the rate $c$ is taken to be the same on all nodes.  Note that individuals are never copied between two nodes belonging to the same group. See Figure~\ref{bipartite}.

To analyse the copying dynamics on this network, we follow the strategy used above for Wright's island model, starting with the single-particle dynamics.  One can verify by inspection that the eigenvalues and eigenvectors of the matrix $\MI$  have the structure presented in Table~\ref{tab:evs}.  Note in particular that the elements of the zero left eigenvector give the stationary distribution $Q_i$ (up to a normalisation constant).  Hence,
\begin{equation}
Q_i = \left\{ \begin{array}{ll}
\displaystyle\frac{m_{BA}}{n_A m_{BA} + n_B m_{AB}} & 1 \le i \le n_A \\[3ex]
\displaystyle\frac{m_{AB}}{n_A m_{BA} + n_B m_{AB}} & n_A < i \le n_A+n_B
\end{array} \right.
\end{equation} 
Meanwhile, the second-smallest eigenvalue is
\begin{equation}
\label{bplrI}
\lrI = \min \{ n_A m_{BA}, n_B m_{AB} \} \;.
\end{equation}

\begin{table}
\begin{center}
\begin{tabular}{lll}
Eigenvalue & Degeneracy & Left eigenvector \\\hline
$\lambda_0 = 0$ & $1$ & $(m_{BA}; m_{AB})$ \\
$\lambda_{1A} = n_B m_{AB}$ & $n_A - 1$ & $( [1,-1]; 0)$ \\
$\lambda_{1B} = n_A m_{BA}$ & $n_B - 1$ & $( 0; [1,-1])$ \\
$\lambda_2 = n_A m_{BA} + n_B m_{BA}$ & $1$ & $(-n_B; n_A)$
\end{tabular}
\end{center}
\caption{\label{tab:evs} Eigenvalues of the single-particle random walk on the bipartite network.  $(x;y)$ denotes a $N$-dimensional vector whose first $n_A$ elements are given by $x$ and the remaining $n_B$ elements by $y$.  The notation $[1,-1]$  indicates a vector of the required dimensionality that has one element equal to $1$, another equal to $-1$ and the rest equal to $0$.}
\end{table}

To bound the decay rate of the quasi-stationary state, $\lqII$, we make use of an extension to the variational ansatz described in the previous section (see Appendix for details).  This extended ansatz still takes the quasi-stationary eigenvector $\phi_{ij} = 1$ when two particles occupy different sites $i\ne j$.  However, it allows for site-dependent values $\phi_{ii} = \alpha_i$ when both particles occupy the same site.  The physical idea behind this ansatz is that the reaction process most strongly affects the probability of finding two particles on the same site, and has a weaker effect elsewhere.

A key quantity in this analysis is the \emph{inverse participation ratio}, which is defined as
\begin{equation}
\label{ipr}
{\cal I} \equiv \sum_i Q_i^2 \;.
\end{equation}
Roughly speaking, ${\cal I}^{-1}$ measures the number of sites on which the particle may be found in the stationary state (see \cite{weg80} for further discussion).  If ${\cal I}$ vanishes as $N\to\infty$, the stationary distribution is delocalised, and $\lqII$ can be shown to be asymptotically bounded from above by
\begin{equation}
\label{default}
\lpd \sim \sum_i Q_i^2 \frac{2 m_i c_i}{2 m_i + c_i}
\end{equation}
where $m_i = \sum_{j\ne i} m_{ij}$ is the total rate at which a particle departs from site $i$, and $c_i$ is the coalescence rate on site $i$.  Furthermore, if $\lpd$ also vanishes with $N$, the expression given above is the optimal bound on $\lqII$ within the space of quasi-stationary eigenvectors admitted by the ansatz. These result are derived the Appendix. We note also that (\ref{default}) coincides with an estimate for the effective size that we obtained previously by more approximate means \cite{bax08}.   

In the case of the bipartite network, $m_i$ takes on two values: $m_i = n_B m_{AB}$ if $i$ is in group $A$ and $m_i = n_A m_{AB}$ if $i$ is in group $B$.  Recalling that $c_i=c$ is constant, we find
\begin{equation}
\label{bilpd}
\lqII \le \lpd \sim \frac{2n_A n_B m_{AB} m_{BA} c}{(n_A m_{BA} + n_B m_{AB})^2} \left[ \frac{m_{BA}}{2n_B m_{AB} +c} + \frac{m_{AB}}{2n_A m_{BA} +c}  \right]
\end{equation}
if the inverse participation ratio vanishes with $N$.  We can use this bound, along with (\ref{bplrI}), in (\ref{cond3}) to establish cases where the diffusion equation (\ref{demf}) is valid for the bipartite network.

In fact, one can obtain $\lqII$ exactly for the bipartite network by following the same procedure as for Wright's island model---that is, by recognising that there are five distinct two-particle configurations in the quasi-stationary state, and identifying each with a variational parameter.  If one follows this path, one finds that when ${\cal I}$ vanishes, so does the ratio (\ref{cond2}), and that $\lqII = \lpd$ asymptotically.  That is, we reach exactly the same conclusions with this more complicated approach, so we present only the results obtained using the simpler analysis just described.  

Of the three rules discussed in Section~\ref{vilsec}, the voter dynamics and invasion process are the most interesting.  We discuss these two cases separately.

\subsubsection{Voter dynamics}

Within voter dynamics, we have
\begin{equation}
\label{bipvot}
m_{AB} = \frac{m}{n_B} \quad\mbox{and}\quad m_{BA} = \frac{m}{n_A} \;.
\end{equation}
The rate of relaxation to the single-particle steady state is simply $\lrI=m$. Meanwhile the inverse participation ratio ${\cal I} = \frac{N}{4n_A n_B}$, which vanishes as long as both $n_A$ \emph{and} $n_B=N-n_A$ grow with $N$ as $N\to\infty$.  When this is the case (\ref{cond3}) is satisfied, and the asymptotic rate of decay of the quasi-stationary state is
\begin{equation}
\label{lastbipv}
\lpd \sim \frac{2mc}{2m + c} {\cal I} = \frac{2mc}{2m + c} \frac{N}{4n_A n_B} \;.
\end{equation}
In fact, since in the standard voter dynamics, the total rate of copying into a site, $m_i$, is the same for every site, $\lpd$ is proportional to the inverse participation ratio on any network. We remark that in the limit $\frac{m}{c} \to 0$ in which the traditional discrete-frequency voter model (see Section~\ref{dcfsec}) is expected to be recovered, the result (\ref{lastbipv}) corresponds with that obtained for the voter model on bipartite networks obtained by other means \cite{soo05,sch09}.

\begin{figure}
\begin{center}
\includegraphics[width=0.66\linewidth]{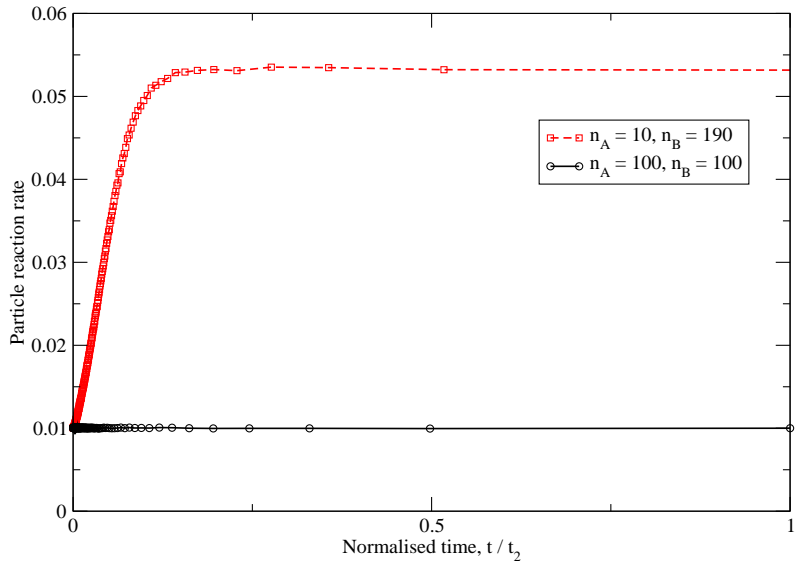}
\end{center}
\caption{\label{bipvotr} Mean reaction rate per particle pair as a function of time for coalescing random walks on bipartite graphs with hop rates chosen to correspond to the voter model and infinite on-site reaction rate. Symbols show the mean reaction rate within an $n$-particle sector plotted at the mean time $t_n$ between entering and exiting that sector. These averages were taken over 100000 realisations of the reaction dynamics on any given network structure.  Time has been rescaled according to $t_2$ to allow direct comparison of two different bipartite network structures: one has equal group sizes whilst in the other they differ greatly.}
\end{figure}

If, say, $n_A$ remains finite as $N\to\infty$, the inverse participation ratio no longer vanishes.  The stationary distribution is localised within the smaller group.  As a consequence, condition (\ref{cond3}) is no longer satisfied, suggesting (but not necessarily implying) that there is no separation of timescales in the two-particle dynamics.  We test this explicitly with Monte Carlo simulations of the backward-time particle reaction dynamics.  Initially, we place one particle on each site, and use the Gillespie algorithm \cite{gil76} to determine the next particle to hop and the time that it does so. If two particles occupy a site simultaneously, they immediately coalesce: thus we are directly simulating the limit $\frac{m}{c} \to 0$.  In Figure~\ref{bipvotr}, we show an empirical measure of the pairwise reaction rate as a function of time.  This is achieved by taking the reciprocal of the mean time spent in the $n$-particle state (averaged over multiple runs), and dividing by ${n \choose 2}$ to convert the total reaction rate to a rate per particle pair.  As can be seen from Figure~\ref{bipvotr}, in the case where $n_A \ll N$, there is a significant transient in which the reaction rate departs from its asymptotic value.

\begin{figure}
\begin{center}
\includegraphics[width=0.66\linewidth]{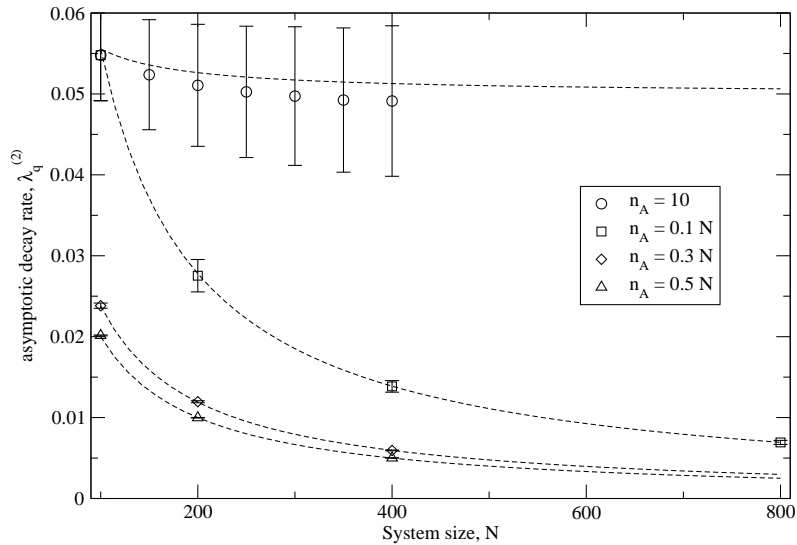}
\end{center}
\caption{\label{bipvotlofN} Mean reaction rate per particle pair measured in computer simulations of coalescing random walks on bipartite graphs with hop rates chosen to correspond to the voter model.  Circles show the case where one group has a fixed, finite size for different values of the total system size $N$; and other symbols cases where both group sizes are proportional to $N$.  Error bars indicate the size of the change in rate over the history of the dynamics (see text for further discussion).  The theoretical predictions for the asymptotic decay rate (\ref{lastbipv}) are shown as dashed lines.  The discrepancies evident for the case of a fixed group size are expected within the theoretical treatment developed here (see text).}
\end{figure}

Behaviour in the limit $N\to\infty$ is revealed by Figure~\ref{bipvotlofN}.  Here, a fit to the asymptotic reaction rate as a function of $N$ is compared for the case $n_A={\rm const}$ versus $n_A \propto N$.  This is done by fitting a constant function to data series like those shown in Figure~\ref{bipvotr}. The error bar indicates the standard deviation of the set of rates: this gives an indication of how much transients are contributing to the average.  As can be seen, when $n_A$ remains constant as $N$ increases, there is no evidence for a decrease in the contribution of the transient, suggesting that the early-time relaxational dynamics cannot be neglected in the limit $N\to\infty$, and thus that the diffusion (\ref{demf}) is indeed not valid in this case.

\begin{figure}
\begin{center}
\includegraphics[width=0.66\linewidth]{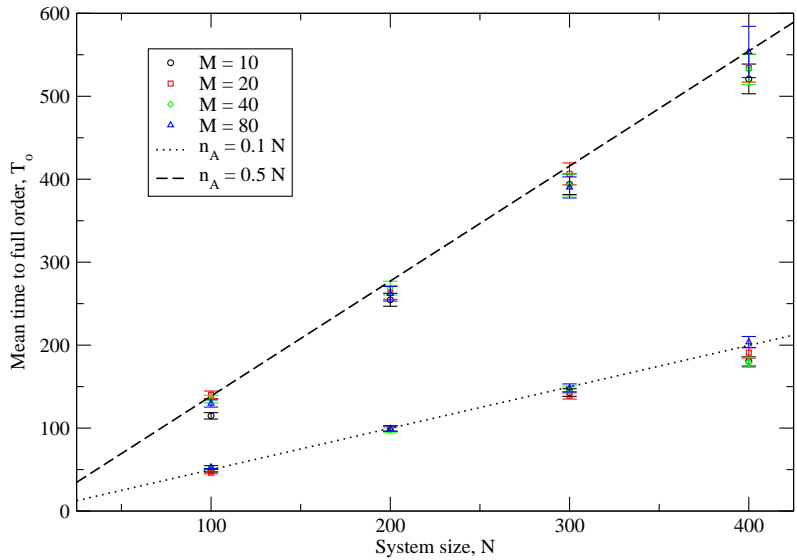}
\end{center}
\caption{\label{bipfvottofN} Mean ordering time for the voter model on the bipartite graph with subpopulation sizes $M=10,20,40$ and $80$ and group sizes $n_A$ and $n_B$ both proportional to the system size $N$.  Each point was obtained from 500 realisations of the forward-time ordering dynamics from an initial condition comprising equal numbers of two species on each site (the exception is the point with the longest ordering time, $M=80$, $n_A=n_B=200$, for which 200 samples were obtained).  The dashed lines show predictions from the $M\to\infty$ theory via Eq.~(\ref{NeMinf}).}
\end{figure}

We also examine the effect of intermediate subpopulation size $M$ (see Section~\ref{modsec} for the precise definition) on ordering times.  We recall that the limit $M\to\infty$ is used to obtain the Fokker-Planck equation (\ref{fpe}). At finite $M$ it is possible that additional terms may contribute.  To investigate this, we performed direct Monte Carlo simulations of the forward-time random-copying dynamics specified in Section~\ref{modsec} with copying rates given by (\ref{muV}).  We took the overall rate of copying between sites $\mu=\frac{1}{M}$, so that then $m_{ij} = \frac{A_{ij}}{z_i}$, $m=1$ and $c_i = c = 2 (1-\frac{1}{M})$.  Initially, half of the $M$ individuals on each site was taken to be of the species of interest, so that the initial value of $\xi=\frac{1}{2}$.  In the limit $M\to\infty$, the effective size appearing in (\ref{demf}) is found from (\ref{lastbipv}) to be
\begin{equation}
\label{NeMinf}
N_e = \frac{8 n_A n_B}{N} \;.
\end{equation}
From a solution of a backward equation corresponding to (\ref{demf}), as described for example in \cite{kim69,soo05}, one finds that the mean time $T_{\rm o}$ for complete order to be reached (i.e, for $\xi$ to reach $0$ or $1$) from the initial condition $\xi=\frac{1}{2}$ is given by $T_{\rm o} = N_e \ln 2$.  The result of performing many repetitions of the forward-time copying dynamics until order is reached, and averaging the ordering time, is shown in Figure~\ref{bipfvottofN} for various $M$.  We see that even with $M=10$, agreement with the $M\to\infty$ result is reasonable. Closer inspection of the data reveals that $T_{\rm o}$ is systematically overestimated in this small $M$ regime; however, already for $M=80$ the data are statistically indistinguishable from the infinite $M$ prediction.  Thus we find that even relatively small subpopulations are reasonably well-described by the continuous frequencies $x_i$.

\subsubsection{Invasion process}  In the invasion process, the hop rates of Eq.~(\ref{bipvot}) are exchanged. The relaxation rate of the single-particle random walk is
\begin{equation}
\lambda_r^{(1)} = m \times \min \left\{ \frac{n_A}{n_B}, \frac{n_B}{n_A} \right\} \;.
\end{equation}
Meanwhile, the inverse participation ratio is given by
\begin{equation}
{\cal I} = \frac{n_A^3 + n_B^3}{(n_A^2 + n_B^2)^2}
\end{equation}
which vanishes as $N\to\infty$ for any combination of $n_A$ and $n_B$, as can be verified by taking $n_A \sim a N^{\gamma}$ with $0 \le \gamma \le 1$ and $n_B=N-n_A$.  Then, (\ref{bilpd}) is asymptotically optimal.  Using (\ref{cond3}), we then find, again for any combination of $n_A$ and $n_B$, that the single-coordinate diffusion (\ref{demf}) applies for the invasion process on the bipartite graph.

\begin{figure}
\begin{center}
\includegraphics[width=0.66\linewidth]{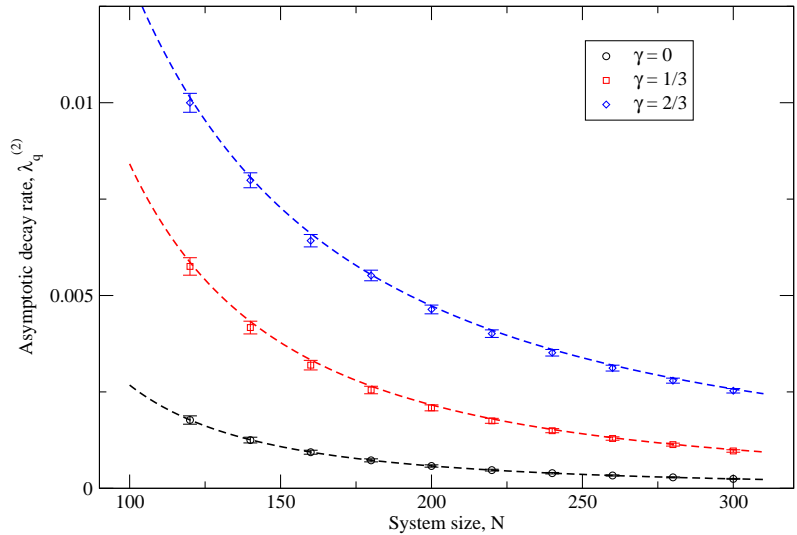}
\end{center}
\caption{\label{bipiplofN} Mean reaction rate per particle pair measured in computer simulations of coalescing random walks on bipartite graphs with hop rates chosen to correspond to the invasion process.  The $A$ group had a size $n_A = a N^\gamma$, and the $B$ group $n_B=N-n_A$.  The choices of $a$ and $\gamma$ used were $(a,\gamma) = (10, 0), (5, \frac{1}{3}), (\frac{3}{2}, \frac{2}{3})$. The dashed lines show the prediction from Eq.~(\ref{lastbiping}). As in Figure~\ref{bipvotlofN}, the error bars reflect the size of the contribution from the transient in the numerical estimate of the asymptotic decay rate of the quasi-stationary state.}
\end{figure}

For $\gamma<1$ in the above group-size definitions, the estimate for the decay rate of the quasi-stationary state takes the particularly simple large-$N$ form
\begin{equation}
\label{lastbiping}
\lpd \sim \frac{2ma}{N^{2-\gamma}} \;.
\end{equation}
We compare the predictions of this formula in the limit $\frac{m}{c}\to0$ against simulations of the type discussed above in the context of voter dynamics.  As can be seen from Figure~\ref{bipiplofN}, numerical measurements of the asymptotic decay rate of the quasi-stationary state agree well with the prediction from Eq.~(\ref{lastbiping}).  Again this provides evidence that the dynamics of discrete-frequency models is well-represented by the $m\to0$ limit of their continuous-frequency counterparts.

We remark that the result (\ref{lastbiping}) \emph{differs} from that previously obtained within an approximation that assumes an absence of degree correlations in the network \cite{soo08}. Thus the invasion process on the bipartite network is a simple, illustrative example of a case where degree correlations must be taken into account to obtain the correct reduction onto a single-coordinate diffusion of the form (\ref{demf}).  Although the correct result was obtained in \cite{sch09}, the separation of timescales required for its validity was assumed, rather than explicitly demonstrated.

\subsection{Correlations in random network structures}

\begin{figure}
\begin{center}
\includegraphics[width=0.66\linewidth]{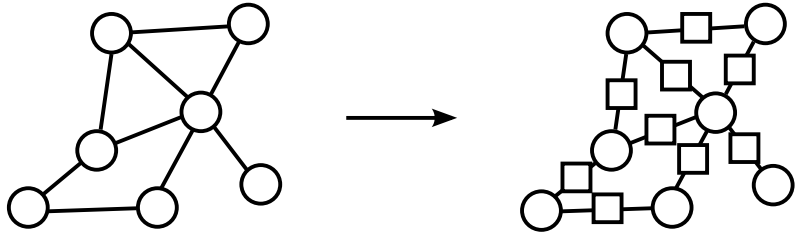}
\end{center}
\caption{\label{explosion} Explosion operation that induces degree anti-correlations in a network.  The exploded network (right) is constructed from the template network (left) by inserting a node (shown as squares) at the mid-point of each edge.}
\end{figure}

We now explore this last point in a little more detail by investigating a transformation to a network structure that induces degree anti-correlations.  We achieve this by beginning with some template network and then `exploding' it by inserting a node in the middle of each edge of the template.  See Figure~\ref{explosion}.  This way, every node of the template network is then next to a node of degree two: thus high-degree nodes end up next to low-degree nodes.

Clearly, one consequence of this transformation is to increase the size of the network, with a likely corresponding increase in the effective size $\Ne$.  Our aim in this final section is to demonstrate that degree anti-correlations lead to a further increase in $\Ne$ beyond what is predicted by (\ref{default}).

As template networks we use Molloy-Reed networks \cite{mol95} with an approximate power-law degree distribution.  These networks are designed to have weak degree correlations.  They are constructed by assigning to each node $i$ its desired degree by attaching $z_i$ half-edges to it.  Then, these half-edges are formed into complete edges by picking a random pair and joining them together if the two nodes in question are not already connected.  Once all half-edges are joined, the probability that nodes $i$ and $j$ are connected is $z_i z_j / {\cal K}_1$ where
\begin{equation}
\label{K}
{\cal K}_m = \sum_i z_i^m \;.
\end{equation}
For this study, we use template networks with the degree distribution
\begin{equation}
\label{pzgamma}
p_z = A_\gamma \int_{z-\frac{1}{2}}^{z+\frac{1}{2}} {\rm d}y\, y^{-\gamma}
\end{equation}
where $A_\gamma$ is a normalisation constant.  For large $z$, this approaches the power-law distribution $p_z \sim z^{-\gamma}$.  We shall however mostly be looking at the behaviour on relatively small networks: the purpose of using the above form for $p_z$ is to allow the possibility of highly heterogeneous networks (i.e., a large variance in the degree $z$).  We shall also, for simplicity, restrict ourselves to voter dynamics, i.e., the case where $m_i=m$ on every node, and the limit $\frac{m}{c}\to0$.  Then, asymptotically, we have from (\ref{lastbipv}) and (\ref{ipr}) an upper bound
\begin{equation}
\label{ranid}
\lpd \sim 2m {\cal I} = 2m \frac{{\cal K}_2}{{\cal K}_1^2}
\end{equation}
on $\lqII$, since $Q_i = \frac{z_i}{{\cal K}_1}$ under voter dynamics \cite{suc05epl,soo05}.  As previously, this bound is valid in the large $N$ limit when ${\cal I}$ vanishes with $N$.

\begin{figure}
\begin{center}
\includegraphics[width=0.66\linewidth]{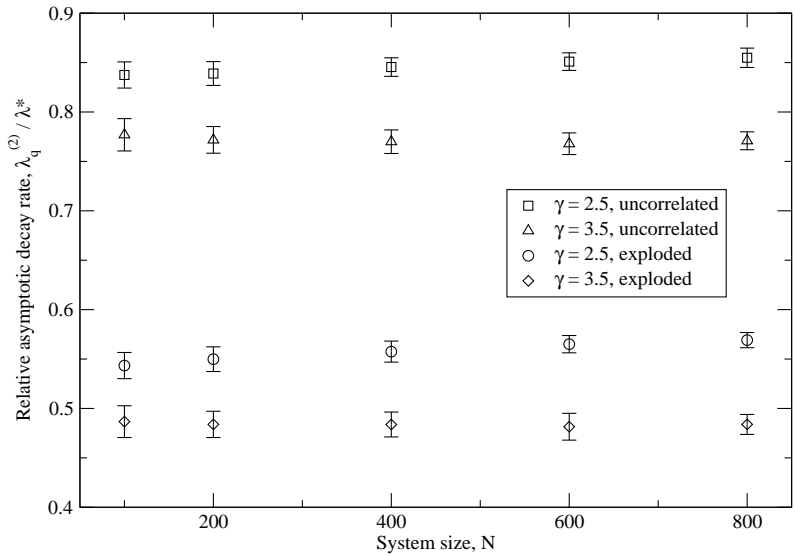}
\end{center}
\caption{\label{ranvotbofN} Empirical measurement of the asymptotic decay rate $\lqII$ relative to the baseline $\lpd$ of Eq.~(\ref{ranid}) as a function of the size of the uncorrelated template network.  Two different values of the exponent $\gamma$ were used in the power-law degree distribution (\ref{pzgamma}).  50 different random template networks were drawn from this distribution for each $\gamma$, and the reaction dynamics repeated 100000 times on both the template and exploded network.  In this figure, error bars indicate the standard error of the distribution of reaction rates within the random network ensembles \emph{not} the standard error of the mean.}
\end{figure}

In Figure~\ref{ranvotbofN} we show empirical measurements of the asymptotic decay rate of the quasi-stationary state, divided by (\ref{ranid}).  We used two degree distributions, one with $\gamma=2.5$, and one with $\gamma=3.5$.  In both cases the minimum degree of any node was $4$.  For the original template we find deviations from this prediction of $10$--$20\%$.  Since (\ref{ranid}) was obtained by assuming an absence of correlations between particles occupying different sites in the quasi-stationary two-particle state, it would appear that (at least at the network sizes simulated) such correlations exist, despite the fact that degrees of neighbouring nodes are designed to be weakly correlated in Molloy-Reed networks.  We also see that the deviations from (\ref{ranid}) increase dramatically when anti-correlations are introduced by the explosion procedure discussed above.  Although we have not been able to demonstrate analytically that there is a separation of timescales in the dynamics on these networks, numerical data (not shown) suggest that the transient is irrelevant except possibly on the uncorrelated template network with $\gamma=2.5$.

To understand more quantitatively the decrease of $\lqII$ (and hence increase of $\Ne$) under the explosion operation, we employ a refinement of the ansatz used to obtain (\ref{ranid}) that allows for correlations between unreacted particles on neighbouring sites in the quasi-stationary state.  Details are presented in the Appendix.  The bound on $\lqII$ that is ultimately obtained in the $\frac{m}{c}\to0$ limit is, for large $N$,
\begin{equation}
\label{ranimp}
\lpr \sim  2m \frac{{\cal K}_2}{{\cal K}_1^2}  \frac{1 - \frac{{\cal K}_2}{{\cal S} - {\cal T}}}{1 - \frac{ \cal S}{{\cal S} - {\cal T}} \frac{{\cal K}_2}{{\cal K}_1^2} } \;,
\end{equation}
where
\begin{eqnarray}
\label{S}
{\cal S} &=& \sum_{ij} z_i A_{ij} z_j \\
\label{T}
{\cal T} &=& \sum_{ijk} z_i A_{ij} A_{jk} A_{ki} \;.
\end{eqnarray}
Here, ${\cal S}$ is a measure of how much high-degree nodes tend to be connected to other high-degree nodes: it is related to the \emph{assortativity coefficient} of a network \cite{new02}.   Meanwhile, ${\cal T}$ comprises a sum over triangles in the network, weighted by the total degree of the vertices. It is thus similar in spirit to a \emph{clustering coefficient} \cite{wat98} in that it vanishes if the network contains no triangles, that is, if locally it has a tree-like structure.  The way these quantities interact in (\ref{ranimp}) is unfortunately not straightforward to interpret.  However, the deviation from the baseline (\ref{ranid}) will typically be large when degrees of neighbouring nodes are anticorrelated and/or triangles tend to be formed by high-degree nodes.

If nodes are uncorrelated, $A_{ij}$ can be replaced with its ensemble average value in the limit $N\to\infty$, that is, with the probability $z_i z_j/{\cal K}_1$ that nodes $i$ and nodes $j$ are connected \cite{soo05}.  In this case, one finds asymptotically that $\lpr$ and $\lpd$ converge.  On the other hand, the bounds are distinct after the explosion operation.  Suppose the template network has given values of ${\cal K}_1$ and ${\cal K}_2$.  Then, after the explosion, there are an additional ${\cal K}/2$ additional nodes of degree two.  This implies that for the exploded network we have ${\cal K}' = 2 {\cal K}$ and ${\cal L}' = {\cal L} + 2 {\cal K}$ (where we use a prime to indicate the exploded network). To compute the statistic ${\cal S}'$, we note that in the exploded graph, each original node $i$ is connected to $z_i$ nodes of degree two.  Taking into account that the sum over all neighbouring pairs $i,j$ can be written as twice the sum over all edges, and that each edge is connected to one of the original template nodes, we find that
\begin{eqnarray}
{\cal S}' = 4 \sum_i z_i^2 = 4 {\cal L} \;.
\end{eqnarray}
By design, the exploded network contains no triangles, so ${\cal T}' = 0$.  Using these values in (\ref{ranimp}) we find, again in the $\frac{m}{c}\to0$ limit,
\begin{equation}
\lpr \sim 2m {\cal I}' \left( \frac{3}{4} - \frac{{\cal K}_1}{2{\cal K}_2} \right) \;,
\end{equation}
which we can contrast with the baseline $\lpd \sim 2m {\cal I}'$.  Relative to this baseline we see that the asymptotic decay rate of the quasi-stationary state is reduced by a factor of at least $\frac{3}{4}$ as a consequence of the degree anti-correlations in the exploded networks.

\begin{figure}
\begin{center}
\includegraphics[width=0.66\linewidth]{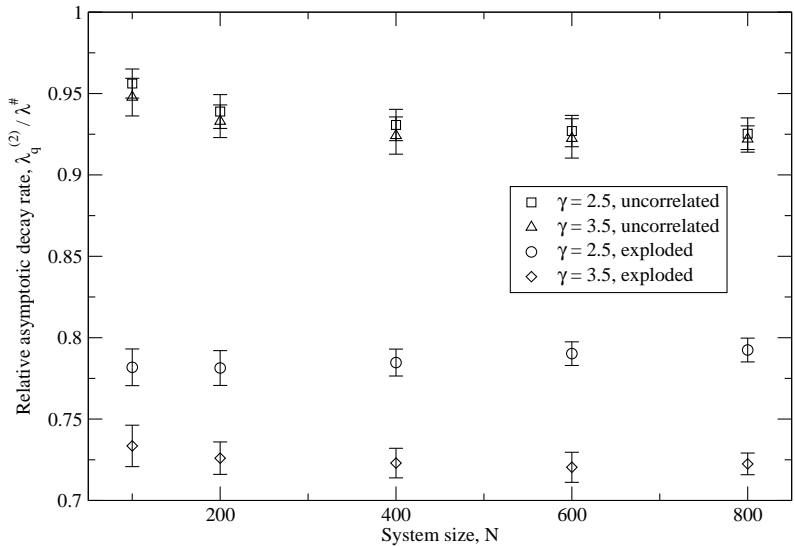}
\end{center}
\caption{\label{ranvotibofN} The asymptotic decay rates of Figure~\ref{ranvotbofN} rescaled relative to the improved estimate $\lambda^{\#}$ of Eq.~(\ref{ranimp}).  Error bars are to be interpreted as in Figure~\ref{ranvotbofN}.}
\end{figure}

In Figure~\ref{ranvotibofN} the same data as in Figure~\ref{ranvotbofN} are shown, but this time relative to the improved bound (\ref{ranimp}).  We see that in all cases, the improved bound is closer to the measured asymptotic decay rate, although significant discrepancies remain.  The reason for this is that the ansatz for the quasi-stationary eigenvector $\phi_{ij}$ does not contain sufficient structure to capture all the relevant correlations between a pair of unreacted particles.  However, the fact that the correction factors shown in Figures~\ref{ranvotibofN} and \ref{ranvotbofN} are only weakly dependent on system size $N$ suggests that the correct \emph{scaling} of the quasi-stationary reaction rate with $N$ is captured by the expressions (\ref{ranid}) and (\ref{ranimp}).  This is consistent with previous works in which the scaling of $\Ne$ with $N$ has been established through fairly simple approximation schemes \cite{vil04,soo05,cas05aip,soo08}, further work having been necessary to improve on the prefactors \cite{bax08,pug09}.  Further improvements demand a more detailed understanding of the structure of the quasi-stationary two-particle state.

\section{Discussion and outlook}
\label{consec}

In this work, we have shown that the onset of order within a large class of random copying processes (including the voter model) on spatially-structured networks can be described exactly in terms of the diffusion of a single collective coordinate $\xi$. This quantity is a weighted average of a species frequency over nodes of the network.  When the reduction onto the single-coordinate diffusion process is possible, the only significant effect of the spatial structure relative to the exactly-solvable `mean-field' diffusion of Equation~(\ref{demf}) is to replace the actual network size $N$ with an effective size $\Ne$.

As we have shown in Section~\ref{redsec}, a remarkable feature of the analysis is that in order to establish the validity of the single-coordinate diffusion (\ref{demf}), one \emph{only} needs to establish two properties of a two-particle reaction-diffusion process.  The first is the lifetime of the quasi-stationary probability distribution for two unreacted particles.  The second is a separation of timescales between this quasi-stationary lifetime and that of the relaxation to quasi-stationarity.  In principle, to establish the probability that complete order has set in by a time $t$, one would need to compute an $N$-point correlation function, starting, for example, with the full $N$-dimensional diffusion equation (\ref{fpe}).  The fact that one can show that the one-dimensional diffusion (\ref{demf}) is valid by appealing only to a two-particle problem represents a considerable simplification in the analysis of voter and other random-copying models.

In Section~\ref{appsec} we demonstrated the separation of timescales within certain exactly-solvable models, namely, Wright's island model and on the bipartite graph.  Although this property has previously been assumed with varying degrees of explicitness within these models \cite{soo08,bax08,sch09}, we are not aware of this property actually being confirmed for these models before.  Furthermore, many existing treatments of random-copying dynamics on heterogeneous networks \cite{soo05,cas05aip,soo08,vaz08,sch09,pug09} have derived a diffusion approximation by aggregating multiple sites in order to construct continuous spatial variables.  Such approaches have the drawback that the network structure gets `washed out' in the coarse-graining.  Here we have described an alternative approach whereby a continuous variable is associated with each site, and the network structure is retained.  As we have shown in Section~\ref{appsec}, anti-correlations between the degree of neighbouring nodes leads in the example of the invasion process on a bipartite network to a scaling of the effective size $\Ne$ with network size $N$ that differs from that predicted by an approach that neglects these correlations \cite{soo08}.  For voter dynamics on random networks, the effect of degree correlations appears to be less severe, affecting the prefactor rather than the scaling exponent.  Since the inverse participation ratio ${\cal I}$ seems to be the quantity that dominates the scaling of $\Ne$ with $N$ in these models, the physical origin of the sublinear scaling of $\Ne$ with $N$ that can be observed \cite{soo05,cas05aip,soo08,sch09,mas10} lies in the localisation of the quasi-stationary single-particle distribution $Q_i$ in the vicinity of high-degree nodes.

In existing treatments of random-copying dynamics on heterogeneous networks, the methods used to arrive at the diffusion equation (\ref{demf}) have to some extent been \textit{ad-hoc}.  Here we have formulated a more systematic treatment based on timescales of the two-particle reaction-diffusion dynamics.  The remaining task is to find good methods to estimate these timescales---and thereby demonstrate their separation.  In cases where the single-particle reaction dynamics is time-reversal symmetric, we have shown that one can use the single-particle relaxation rate as a proxy for the corresponding quantity in the two-particle sector; we have also demonstrated the utility of a variational approach to bound the effective size $\Ne$ from below.  While this has allowed the separation of timescales to be demonstrated within very simple models, it was still necessary to resort to simulations to establish whether it was present on random network structures.  Development of these methods---perhaps using bounds on the single-particle relaxation dynamics that have been derived elsewhere \cite{dia91}---would be needed to determine the class of networks and dynamical processes for which (\ref{demf}) is valid.

We remark that (\ref{demf}) is known to apply to the voter model on square lattices in two dimensions and above \cite{cox89}.  In two dimensions, it is found that the effective population size grows logarithmically with system size (relative to the rate at which a particle relaxes to the stationary uniform distribution).  This can be understood as a statement that domains coarsen logarithmically in the two-dimensional voter model \cite{cas09}.  Although the separation of timescales is weak in this case, it is nonetheless present.  Thus one would expect the separation of timescales to be observed on random networks with small diameters, on the grounds that a single particle would be expected to explore the system on a faster timescale than would be possible on a two-dimensional square lattice of the same size.  It would be interesting to pursue this line of enquiry further.

As mentioned in the introduction, the use of a single effective size $\Ne$ to characterise purely stochastic population dynamics in subdivided populations is widespread in population genetics (see e.g., \cite{wan99} for a review).  A fair amount of discussion has centred around the fact that different definitions of $\Ne$ are possible, depending on which property of the stochastic dynamics is measured and calibrated against an unstructured (mean-field) population \cite{whi97}.  Whilst it is widely understood that asymptotically many definitions converge onto the lifetime of the quasi-stationary state (also known as the \emph{eigenvalue effective size} \cite{ewe04}), the importance of a separation of timescales between this lifetime and others seems to be less well established.  For example, Sj\"{o}din \textit{et al} \cite{sjo05}, identify the asymptotic rate of particle coalescence as defining `the' effective population size.  In these analyses, they examine a separation of timescales in the \emph{definition} of migration and coalescence rates, $m_{ij}$ and $c_i$.  That is, migration rates that vanish with local population size $M$ with differing exponents.  This approach potentially misses the possibility---demonstrated here in the examples of Section~\ref{appsec}---that one can have a separation of \emph{emergent} timescales that is sufficient to allow a single effective size that characterises almost the entire history of the dynamics.  The importance of emergent timescales has perhaps been most clearly recognised by Wakeley and co-workers \cite{wak01,wak04,mat06}.  In particular, the rapid relaxation of the single-particle dynamics as a condition for a single, well-defined effective size was noted in \cite{mat06}.  However, that study was restricted to a rather special class of \emph{vertex-transitive} networks and dynamical update rules. Furthermore, the irrelevance of \emph{all} timescales other than the relaxation of the quasi-stationary state was implicitly assumed rather than explicitly demonstrated.  We believe that the present analysis offers considerable additional insight into the applicability of a one-dimensional diffusion characterised by a single effective population size for a general class of random-copying dynamics on complex networks.

Nevertheless, a number of open questions remain.  As we have already noted, the task of calculating the relevant decay rates $\lqII$ and $\lrII$ is in general challenging (despite being simpler than solving the full $N$-body dynamics).  Although we have shown that certain simplifications arise when the single-particle random walk dynamics are time-reversal symmetric, there is of course a much larger class of processes that do not respect this symmetry: these have nonequilibrium steady states.  Here we have not been able to show a result analogous to that derived in Section~\ref{1vs2sec}, that is, to show that the relaxation rate to two-particle quasi-stationarity exceeds the relaxation rate to a single-particle nonequilibrium steady state, although intuitively one might expect this to be the case.  Furthermore, we do not know of any convenient means to bound the asymptotic reaction rate in these models.

Although we have found that we recover known results for discrete-frequency models, or obtain good agreement with simulations, in the limit $\frac{m}{c}\to0$, as we noted in Section~\ref{dcfsec}, this correspondence relies on the limits of infinite local subpopulation size and vanishing migration rate to commute.  Again, we are not aware of a proof that this should be the case.    At the end of Section~\ref{redsec} we commented that the initial condition on the diffusion equation may be modified as the system evolves into the quasi-stationary state: further clarity here would be beneficial.  Additionally, although the coordinate $\xi$ is sufficient to track the degree of order present in a system, other aspects of the evolution cannot be determined from this quantity alone---for example, the typical state of the system or correlation length given some value of $\xi$.  It is possible that some of these quantities can be obtained with knowledge of the quasi-stationary eigenvector $\phi_{ij}$, this by definition encapsulates two-point correlations.  It is unclear whether higher-point correlation functions could be inferred from the two-point functions, or whether these would need to be determined separately.

Finally, we revisit the fundamental property shared by the entire set of random-copying processes discussed here, namely, that all species are equivalent.  There is considerable interest in those cases where one species is fitter than another.  Although in some sense this is only a minor modification to the dynamics, the mathematics become much more involved.  In particular, the backward-time dynamics are more complicated: the reaction rates then depend on the frequencies of different species in the population at the point in history at which the coalescence reaction takes place.  It may, however, be possible to apply the results obtained here for the neutral theory in some kind of perturbation theory to study the effects of a small selective advantage. We leave this as one of many possibilities for future work.

\section*{Acknowledgments}

I thank the RCUK for the support of an Academic Fellowship, Alan McKane and Gareth Baxter for useful discussions during the early development of this work and all those colleagues who have commented on preliminary accounts presented at conferences and research seminars.

\appendix

\section{Convergence to discrete-frequency models in the limit of slow copying rates}
\label{conv}

In Section~\ref{dcfsec} it was noted that a convergence theorem due to M\"{o}hle \cite{moh98} implies that if one takes the between-site copying rates $\mu_{ij} \to 0$ $\forall i\ne j$ at any fixed subpopulation size $M$, the discrete-frequency ($M=1$) dynamics is recovered.  In this appendix we provide an informal account of this mathematical result that illustrates how this limit is reached.  We refer the reader to \cite{moh98} for the technical detail.

Our starting point is a \emph{discrete-time} version of the random-copying process. At any point in time, the state of the system, $\sigma$, is specified by the set of numbers $n_i$ which count the number of individuals of the species of interest on site $i$. In any one timestep, one of these numbers may change by $\pm 1$. The idea now is to split the matrix $A$ of transition probabilities that appears in the master equation $P_{\sigma'}(t+\delta t) = \sum_\sigma A_{\sigma', \sigma} P_{\sigma}(t)$ into two pieces.  The first, $L$, accounts for local interactions: that is, when an individual is replaced by sampling from the same site. The second, $I$, accounts for interactions between sites.  Both are stochastic matrices. If the latter happen a fraction $\epsilon$ of the time, then
\begin{equation}
A = (1-\epsilon) L + \epsilon I \;.
\end{equation}

The theorem in \cite{moh98} then states that the vector of probabilities $P(t)$ has the limiting form
\begin{equation}
\label{mm}
P(n \delta t) \to {\rm e}^{\epsilon n {\cal P} I {\cal P}} {\cal P} P(0)
\end{equation}
as $\epsilon\to 0$, subject to the existence of the projector
\begin{equation}
{\cal P} = \lim_{n \to \infty} L^n \;.
\end{equation}

The way to read (\ref{mm}) is as follows. The projector ${\cal P}$ performs an infinite number of local copying events on every site. If on site $i$ there are initially $n_i$ individuals of the species of interest, then at the end of this process $n_i \to M$ with probability $\frac{n_i}{M}$, and $n_i \to 0$ with probability $1-\frac{n_i}{M}$.  Thus the initial condition is projected onto a state in which every site is occupied by individuals of a single species.  The exponential term in (\ref{mm}) is the generator of a \emph{continuous-time} process. If we take the limit $\delta t \to 0$, a copy between sites is taken at rate $\epsilon$ per timestep. A single copy may leave the system unchanged; or it may change one of the $n_i$ values by $\pm 1$. In the latter case, all but one of the individuals are of the same species. Then, the projector acts, either to restore the existing value of $n_i$ (with probability $1 - \frac{1}{M}$) or to `flip' it (i.e., from $M\to 0$ or vice versa) with probability $\frac{1}{M}$.

Putting this together, one finds that the frequency $x_i=\frac{n_i}{M}$ flips from $0$ to $1$ (or vice versa) due to a copy taken from a neighbouring site $j$ at a rate $\frac{\epsilon}{M}$ multiplied by the probability of copying from $j$ to $i$ in the original discrete-time process.  In Section~\ref{modsec}, the random-copying process was set up as a continuous-time process in which copies from $j$ arriving at $i$ at a rate $\mu_{ij}$.  This is obtained here by putting the transition probabilities encoded in $I$ equal to $\mu_{ij} \delta t$.  Then, in the original time units, a copy between sites is taken at rate $\mu_{ij}' = \frac{\epsilon}{M} \mu_{ij}$.  Rescaling time such that $t' = \frac{\epsilon}{M} t$, we obtain in the limit $\epsilon\to0$ the $M=1$ (discrete frequency) dynamics from any model with $M>1$.  Note that in the case of the dynamical rules (\ref{muV})--(\ref{muL}) we can identify $\epsilon$ with $\mu$.

\section{Variational bounds on the quasi-stationary decay rate}

As stated in the main text, when the backward-time hopping process is time-reversal symmetric, all eigenvalues of the matrix $\MII$ are real, and hence the smallest eigenvalue satisfies the bound
\begin{equation}
\label{vara}
\lqII \le \lup = \frac{\sum_{ijk\ell} Q_i Q_j \phi_{ij} \MII_{ij,k\ell} \phi_{k\ell}}{\sum_{ij} Q_i Q_j \phi_{ij}^2} \;,
\end{equation}
for any vector $\phi_{ij}$.  In order to construct bounds on $\lqII$ one can use a variational approach.   That is, one can include various parameters in $\phi_{ij}$, and minimise the right-hand side of (\ref{vara}), thereby obtaining the most stringent bound on $\lqII$ available within the space of vectors spanned by these parameters.

The key to designing a variational ansatz is to recall that the quasi-stationary distribution of two unreacted particles $Q_{ij} \propto Q_i Q_j \phi_{ij}$.  Thus $\phi_{ij}$ measures the extent to which the probability of finding a pair of particles on sites $i$ and $j$ is affected by the interaction between them. Alternatively, one can infer from the interpretation of $P_{ij} \propto \phi_{ij}$ given at the end of Section~\ref{redsec} that $\phi_{ij}$ is the probability that two particles starting on sites $i$ and $j$ do not react before the onset of the quasi-stationary state.   Either way anticipate that $\phi_{ij}$ will be close to unity for particles that are far apart, and decrease as their separation decreases.

We make use of two variational ans\"{a}tze in the main text.  In the first it is assumed that $\phi_{ij}$ differs from unity only when both particles are on the same site.  In the second we extend to include the effects of interactions when particles occupy neighbouring sites (but at the expense of restricting to voter-type update rules only).

\paragraph{Ansatz invoking on-site interactions only} Restricting the effect of interactions to particles occupying the same site, corresponds to an ansatz for $\phi_{ij}$ of the form $\phi_{ii} = \alpha_i$ and $\phi_{ij} = 1$ if $i \ne j$.  Differentiating (\ref{vara}) with respect to $\alpha_m$, and setting it to zero, we find that
\begin{equation}
\label{on1}
\frac{\sum_{k\ell} \MII_{mm,k\ell} \phi_{k\ell}}{ \phi_{mm} } = \frac{\sum_{ijk\ell} Q_i Q_j \phi_{ij} \MII_{ij,k\ell} \phi_{k\ell}}{\sum_{ij} Q_i Q_j \phi_{ij}^2} = \lup \;.
\end{equation}
Using the definition of $\MII$, Eq.~(\ref{MII}), we find that
\begin{equation}
\sum_{k\ell} \MII_{mm,k\ell} \phi_{k\ell} = 2m_i(\alpha_i-1) + c_i \alpha_i
\end{equation}
where $m_i = \sum_{j \ne i} m_{ij}$, the total exit rate from site $i$ in the random-walk picture.  Substituting into (\ref{on1}), we find that
\begin{equation}
\label{onsim1}
\alpha_i = \frac{2m_i}{2m_i + c_i - \lup} \;.
\end{equation}
This is not sufficient to determine the optimal $\alpha_i$, since the optimal value of $\lup$ is not known.  To compute this, we need an additional relation between them.  This can be obtained as follows. First, we note that (\ref{on1}) can be written as
\begin{equation}
\label{on2}
Q_i^2 \sum_{k\ell} \MII_{ii, k\ell} \phi_{k\ell} = \lup Q_i^2 \phi_{ii} \;.
\end{equation}
Meanwhile, if we temporarily set $\phi_{ij} = \alpha_{N+1}$ for $i\ne j$, differentiate with respect to this additional $\alpha$ parameter and set to zero, we find analogously that
\begin{equation}
\sum_{i \ne j} Q_i Q_j \sum_{k\ell} \MII_{ij, k\ell} \phi_{k\ell} = \lup \sum_{i\ne j} Q_i Q_j \phi_{ij} \;.
\end{equation}
Summing (\ref{on2}) over all $i$, and adding to this expression, we find that
\begin{equation}
\lup = \frac{\sum_{ijk\ell} Q_i Q_j \MII_{ij,k\ell} \phi_{k\ell}}{ \sum_{ij} Q_i Q_j \phi_{ij} } = \frac{\sum_{i} Q_i^2 c_i \phi_{ii}}{\sum_{ij} Q_i Q_j \phi_{ij}}
\end{equation}
where the second equality holds because $Q_i$ is the zero left eigenvector of the matrix $\MI$.  Using the definition (\ref{MII}) and the form of the ansatz for $\phi_{ij}$, we ultimately obtain
\begin{equation}
\label{onsim2}
\lup = \frac{\sum_i Q_i^2 c_i \alpha_i}{1 - \sum_i Q_i^2 (1-\alpha_i)} \;.
\end{equation}
The optimal bound on $\lqII$ is found by solving the simultaneous set of $N+1$ equations given by (\ref{onsim1}) and (\ref{onsim2}).

For Wright's island model, discussed in Section~\ref{wimsec}, $m_i$, $c_i$ and $\alpha_i$ are all site-independent quantities, and $Q_i=\frac{1}{N}$.  Making these substitutions in (\ref{onsim1}) and (\ref{onsim2}), we arrive at the pair of equations (\ref{wimsim1}) and (\ref{wimsim2}) studied in the main text.

For more general models, solving (\ref{onsim1}) and (\ref{onsim2}) exactly is difficult.  Nevertheless, one can still make some progress here.  An alternative way to formulate this problem is to define
\begin{eqnarray}
y(x) &=& \frac{1}{1 - \sum_i Q_i^2} \sum_i Q_i^2 \frac{2m_i(c_i - x)}{2m_i + c_i - x} \;.
\end{eqnarray}
Then, the optimal bound $\lup$ is given by the value of $x$ for which $y(x) = x$. Differentiating, we find
\begin{equation}
\frac{{\rm d} y}{{\rm d} x} = - \frac{1}{1 - \sum_i Q_i^2} \sum_i Q_i^2 \left( \frac{2m_i}{2 m_i + c_i - x} \right)^2 \le 0 \;.
\end{equation}
Hence, the root of $y(x) = x$ must lie at a value of $y(x) \le y(0)$; hence $\lup \le y(0)$.  Thus, one has that
\begin{equation}
\lup \le \frac{1}{1 - \sum_i Q_i^2} \sum_i Q_i^2 \frac{2 m_i c_i}{2m_i + c_i} \;.
\end{equation}
Note that the right-hand side of this expression is not necessarily the optimal bound.

However, if the inverse participation ratio ${\cal I} = \sum_i Q_i^2$ vanishes as $N\to\infty$, then asymptotically we obtain the simpler expression
\begin{equation}
\label{lupa}
\lup \le \lpd \sim \sum_i Q_i^2 \frac{2 m_i c_i}{2m_i + c_i}
\end{equation}
which was used in the main text.  This bound \emph{is} optimal within the present ansatz for $\phi_{ij}$ if $\lup$ also vanishes as $N\to\infty$.  Then, since
\begin{equation}
\lup = \frac{\sum_i Q_i^2 (c_i - \lup) \alpha_i}{1 - \sum_i Q_i^2}
\end{equation}
and $\alpha_i \ge \frac{2m_i}{2m_i + c_i}$,
\begin{equation}
\lup \ge \frac{1}{1-\sum_i Q_i^2} \sum_i Q_i^2 (c_i - \lpd) \frac{2m_i}{2m_i+c_i} \;.
\end{equation}
If both ${\cal I} = \sum_i Q_i^2$ and $\lpd$ vanish as $N\to\infty$, this lower bound on the optimal $\lup$ asymptotically approaches the upper bound (\ref{lupa}), and hence the right-hand side of (\ref{lupa}) is optimal in this case.

\paragraph{Ansatz invoking on-site and nearest-neighbour interactions} As noted in the main text, the traditional voter model dynamics is obtained by taking $m_i=m$ and $c_i=c$ followed by the limit $\frac{m}{c}\to0$.  In the above ansatz, this leads to all $\alpha_i = \frac{2m}{2m+c}$ and $\lpd \propto {\cal I}$.  To examine the effects of degree correlations between neighbouring nodes, it is natural to introduce an ansatz for $\phi_{ij}$ that takes one value when both particles are on the same site, a different value when they are on neighbouring sites, and unity elsewhere.  That is,
\begin{equation}
\phi_{ij} = 1 + (\alpha - 1) \delta_{ij} + (\beta - 1) A_{ij}
\end{equation}
where $A_{ij}$ is the adjacency matrix for the network.  Recalling that $A_{ij} = 0$ if $i=j$, we see that this expression takes the value $\alpha$ if $i=j$, $\beta$ if $i$ and $j$ are neighbours, and unity otherwise. For voter dynamics, we have from (\ref{muV}) and (\ref{MII}) that
\begin{equation}
\MII_{ij,k\ell} = \left(m \delta_{ik} - \frac{m}{z_i} A_{ik}\right) \delta_{j\ell}  + \left( m \delta_{j\ell} - \frac{m}{z_j} A_{j\ell} \right) \delta_{ik} + c \delta_{ijk\ell}
\end{equation}
where $z_i$ is the degree of node $i$.

Differentiating the right-hand side of (\ref{vara}) with respect to $\alpha$ and setting to zero, we find
\begin{eqnarray}
\hspace{15ex} \sum_{ik\ell} Q_i^2 \MII_{ii,k\ell} \phi_{k\ell} &=& \lup \sum_i Q_i^2 \phi_{ii} \\
\implies
\sum_i Q_i^2 \left[ (2m+c) \alpha - 2m \beta \right] &=& \lup \sum_i Q_i^2 \alpha \;.
\end{eqnarray}
Hence, 
\begin{equation}
\label{alnn}
\alpha = \frac{2m \beta}{2m + c - \lup} \;.
\end{equation}
We notice the similarity with (\ref{onsim1}): the on-site correlation $\phi_{ii}$ is multiplied by the nearest-neighbour correlation $\beta$.  To obtain an expression for this correlation, we differentiate the right-hand side of (\ref{vara}) with respect to $\beta$, and set to zero.  This yields
\begin{equation}
\sum_{ijk\ell} Q_i A_{ij} Q_j \MII_{ij,k\ell} \phi_{k\ell} = \lup \sum_{ij} Q_i A_{ij} Q_j \phi_{ij} \;.
\end{equation}

This expression can be evaluated by using the fact that, for the voter model, the stationary probabilities are $Q_i = z_i / {\cal K}_1$.  The sum on the right-hand side becomes
\begin{equation}
\frac{1}{{\cal K}_1^2} \sum_{ij} z_i A_{ij} z_j \phi_{ij} = \beta \frac{{\cal S}}{{\cal K}_1^2} 
\end{equation}
where ${\cal K}_n$ and ${\cal S}$ are defined by Equations (\ref{K}) and (\ref{S}) respectively.  The sum on the left requires a bit more work.  One finds it can be written as
\begin{eqnarray}
\frac{2m}{{\cal K}_1^2} \left( \sum_{ij} z_i A_{ij} z_j \phi_{ij} -  \sum_{ijk }z_i A_{ij} z_j  \frac{A_{ik}}{z_i}\phi_{jk} \right) \nonumber\\
\qquad = \frac{2m}{{\cal K}_1^2} \left(  \beta {\cal S} - \sum_{ijk} z_i A_{ij} A_{jk} \left[ 1 + (\alpha-1) \delta_{ki} + (\beta-1) A_{ki} \right] \right) \nonumber\\
\qquad = \frac{2m}{{\cal K}_1^2} \left( \beta {\cal S}  - {\cal S} - (\alpha-1) {\cal K}_2 - (\beta -1) {\cal T} \right)
\end{eqnarray}
in which ${\cal T}$ is given by (\ref{T}). Hence,
\begin{equation}
2m  \left[ (\beta - 1) ( {\cal S} - {\cal T} ) - (\alpha - 1) {\cal K}_2 \right] = \lup \beta {\cal S}
\end{equation}
Substituting (\ref{alnn}), and rearranging, we finally find
\begin{equation}
\label{benn}
\beta = \frac{{\cal S} - {\cal K}_2 - {\cal T}}{\left( 1 - \frac{\lup}{2m} \right) {\cal S} - \frac{2m}{2m+c - \lup} {\cal K}_2 - {\cal T}} \;.
\end{equation}
Similar manipulations allow one to determine that
\begin{equation}
\label{lnn}
\lup = \frac{\sum_i z_i^2 \phi_{ii} c}{\sum_{ij} z_i z_j \phi_{ij}} = \frac{\alpha c {\cal K}_2}{{\cal K}_1^2 + (\alpha-1) {\cal K}_2 + (\beta - 1) {\cal S}} \;.
\end{equation}

When the inverse participation ratio ${\cal I}$ vanishes with $N$, we know that $\lup$ does too, since (\ref{lupa}) bounds it from above, and is proportional to ${\cal I}$ under voter dynamics.  The upshot of this is that in this case, we can neglect the terms in $\lup$ in (\ref{alnn}) and (\ref{benn}).  There are further simplifications in the limit $\frac{m}{c}\to0$ which is case studied in the main text.  To leading order in $\frac{m}{c}$ one finds
\begin{equation}
\alpha = \frac{2m \beta}{c} \quad\mbox{and}\quad \beta = 1 - \frac{{\cal K}_2}{{\cal S} - {\cal T}} \;.
\end{equation}
Using these in (\ref{lnn}), and further taking into account that the ratio ${\cal I} = \frac{{\cal K}_2}{{\cal K}_1^2}$ is assumed to vanish as $N\to\infty$, we ultimately obtain the expression (\ref{ranimp}) quoted in the main text.

\end{document}